\documentclass[11pt,a4paper]{article}
\pdfoutput=1
\usepackage{jheppub}

\usepackage{color}
\usepackage{amsmath}
\usepackage{pifont}   
\usepackage{verbatim}       
\usepackage{acronym} 
    
\usepackage{amsfonts}
\usepackage{amssymb}
\usepackage{mathrsfs} 
\usepackage{graphicx}
\usepackage{multirow} 
\usepackage{slashed} 
\usepackage{array}

\usepackage{graphicx}
\usepackage{epsfig}
\usepackage{lscape}
\usepackage{rotating}
\usepackage{epsf}
\usepackage{bm}
\usepackage{booktabs}
\usepackage{array}
\usepackage{caption}
\usepackage{subcaption}
\usepackage[utf8]{inputenc}
\usepackage{float}

\usepackage{multirow}
\usepackage{array,booktabs,colortbl,multirow}
\usepackage{colortbl,xcolor,colordvi,color}
\usepackage{rotating}
\allowdisplaybreaks

\newcommand{\cmrule}{\midrule[0.25mm]}

\title{Searching for periodic signals in kinematic distributions using continuous wavelet transforms}

\author{Hugues Beauchesne}
\author{and Yevgeny Kats}

\affiliation{Department of Physics, Ben-Gurion University, \\Beer-Sheva 8410501, Israel}

\emailAdd{beauches@post.bgu.ac.il, katsye@bgu.ac.il}

\abstract{Many models of physics beyond the Standard Model include towers of particles whose masses follow an approximately periodic pattern with little spacing between them. These resonances might be too weak to detect individually, but could be discovered as a group by looking for periodic signals in kinematic distributions. The continuous wavelet transform, which indicates how much a given frequency is present in a signal at a given time, is an ideal tool for this. In this paper, we present a series of methods through which continuous wavelet transforms can be used to discover periodic signals in kinematic distributions. Some of these methods are based on a simple test statistic, while others make use of machine learning techniques. Some of the methods are meant to be used with a particular model in mind, while others are model-independent. We find that continuous wavelet transforms can give bounds comparable to current searches and, in some cases, be sensitive to signals that would go undetected by standard experimental strategies.}

\begin{document}

\maketitle

\section{Introduction}\label{Sec:Introduction}
Experimental searches for physics beyond the Standard Model often look for peaks or dips in kinematic distributions. Although this is certainly well motivated, there exist potential signals that can take far more complicated forms and which have received very little attention.

\begin{figure}[t]
  \begin{center}
    \includegraphics[width=0.95\textwidth]{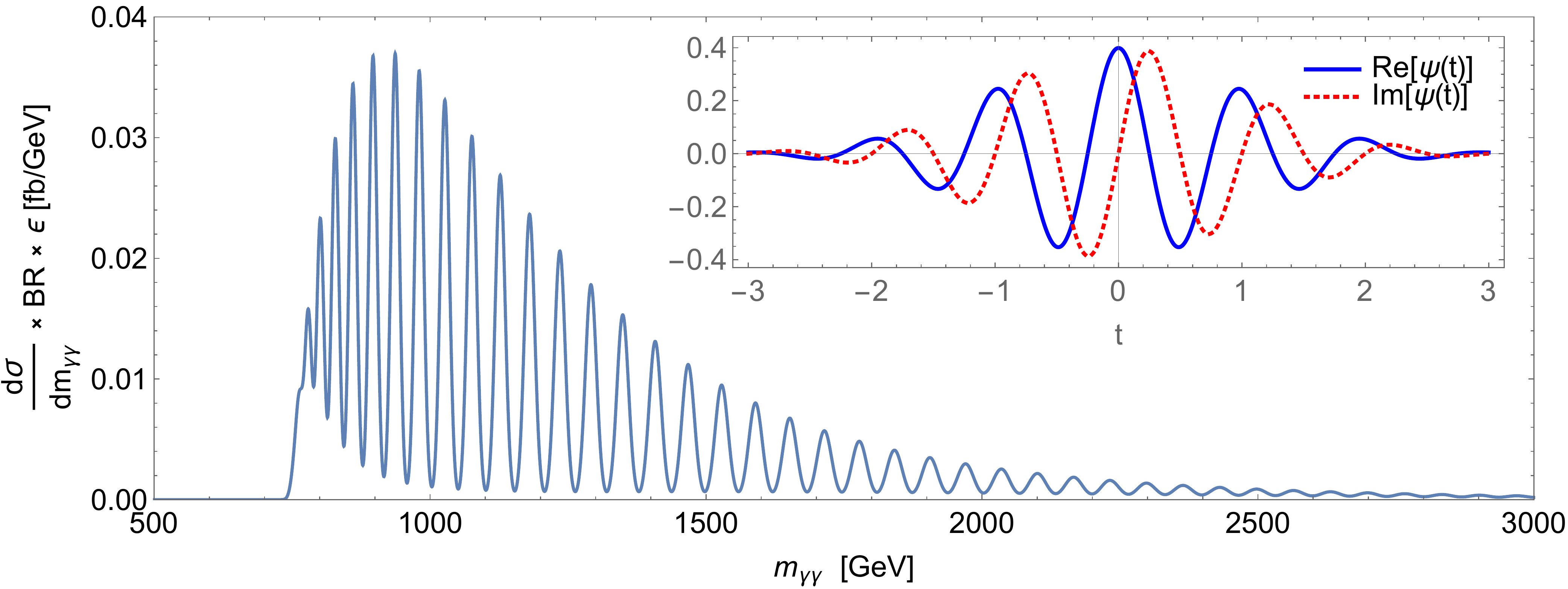}
  \end{center} 
  \caption{Example of a periodic signal in the diphoton spectrum from the clockwork/linear dilaton scenario. The parameters are set to $k = 750$~GeV and $M_5 = 3$~TeV. See appendix~\ref{app:App} for the technical details of the model and the modeling of the experimental resolution. The inset shows the Morlet wavelet (see eq.~(\ref{eq:MorletDef})).}\label{fig:Signal}
\end{figure}

One such possibility is periodic signals. These are a typical signature of models that include a large number of similar resonances with small mass splitting. Such models include the linear dilaton scenario~\cite{Antoniadis:2011qw,Baryakhtar:2012wj,Giudice:2016yja,Giudice:2017fmj}, discrete and continuum clockwork models~\cite{Giudice:2016yja,Giudice:2017fmj}, certain limits~\cite{Giudice:2004mg,Kisselev:2008xv,Franceschini:2011wr} of the Randall-Sundrum model~\cite{Randall:1999ee} and more exotic warped extra dimensions~\cite{Brax:2019koq}. An example of the two-photon invariant mass distribution is shown in figure~\ref{fig:Signal} for the clockwork/linear dilaton (CW/LD) scenario~\cite{Giudice:2017fmj}.

Intuitively, one may think that taking the Fourier transform of a distribution like this would be an ideal strategy to exploit its periodic nature. There are however some complications with this in practice, as in most scenarios signals do not repeat themselves perfectly and indefinitely. For example, the repetitions might only occur over a finite interval. The position of that interval, which could potentially be used to discriminate signal from background, is essentially lost when passing to frequency space. Also, the frequency of the signal may not be constant. The Fourier transform would then not convey clearly which frequency is present at which point. All in all, certain characteristics of realistic periodic signals are easier to see in the time domain, while others are more clear in the frequency domain. As such, neither the signal itself nor its Fourier transform is ideal to discover a signal whose periodicity changes with time.

\begin{figure}[t!]
  \centering
   \captionsetup{justification=centering}
    \begin{subfigure}{0.47\textwidth}
    \centering
    \includegraphics[width=\textwidth]{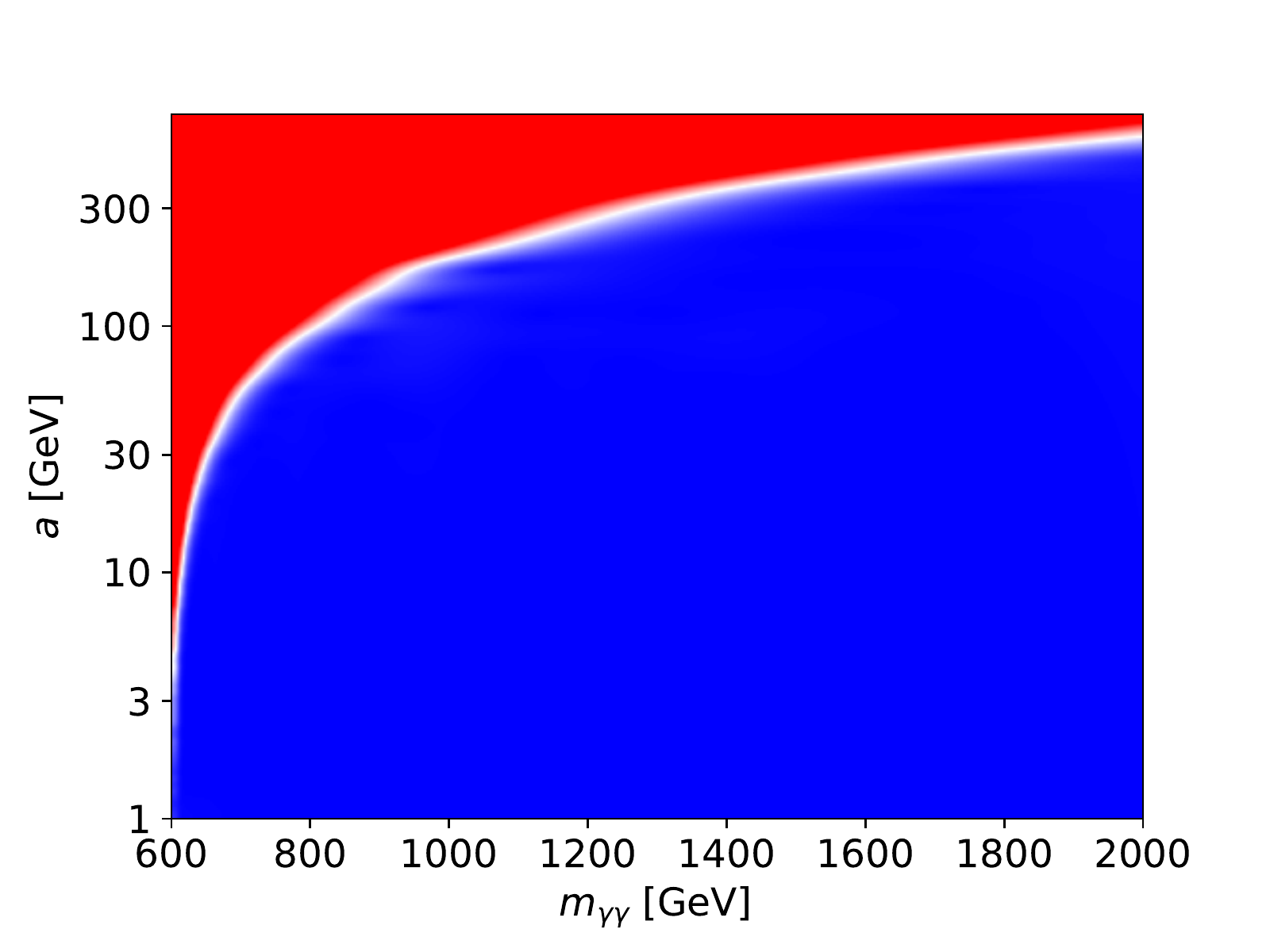}
    \caption{}
    \label{fig:BFix}
  \end{subfigure}
  \begin{subfigure}{0.47\textwidth}
    \centering
    \includegraphics[width=\textwidth]{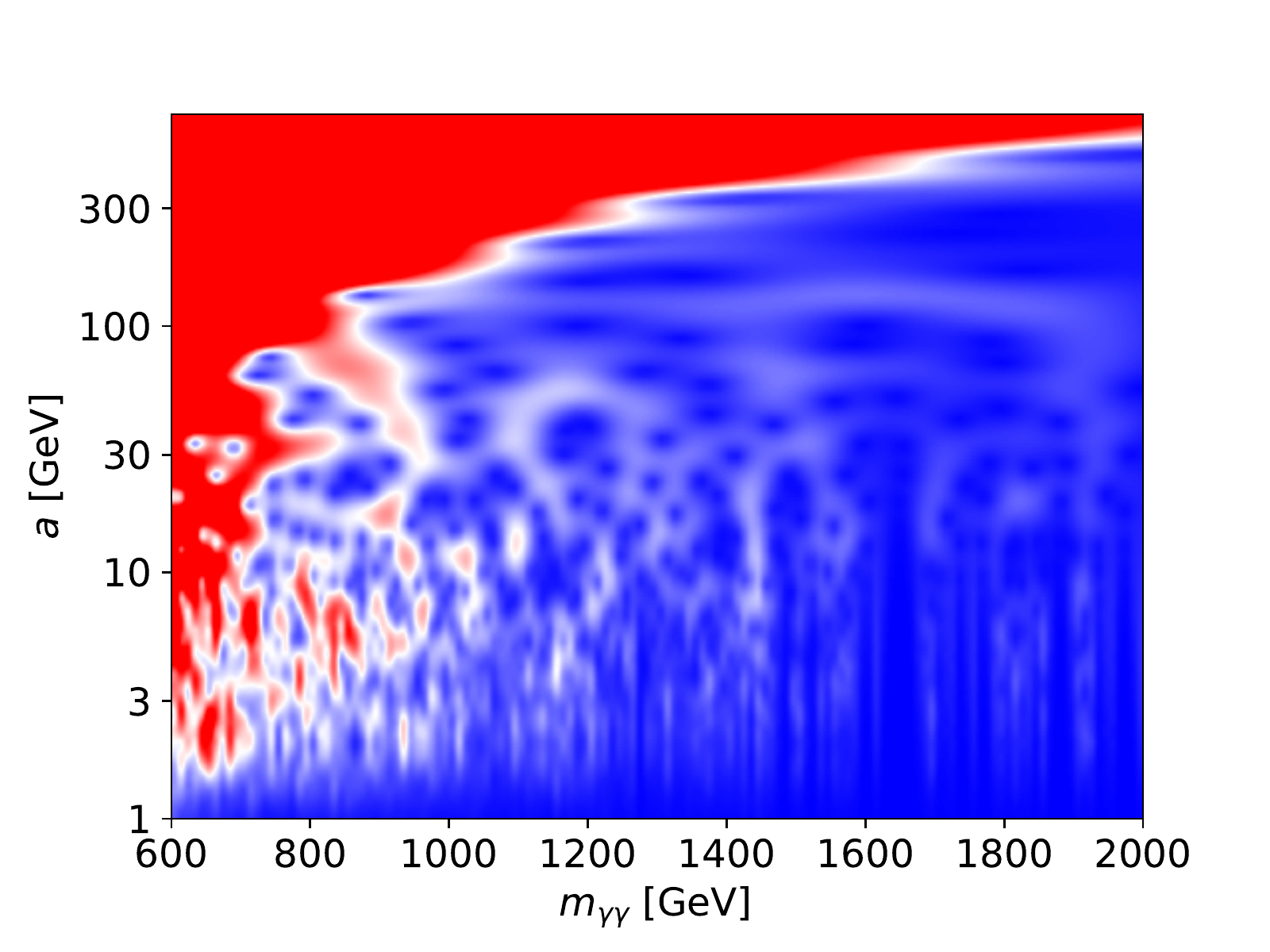}
    \caption{}
    \label{fig:RandomB}
  \end{subfigure}
  \begin{subfigure}{0.47\textwidth}
    \centering
    \includegraphics[width=\textwidth]{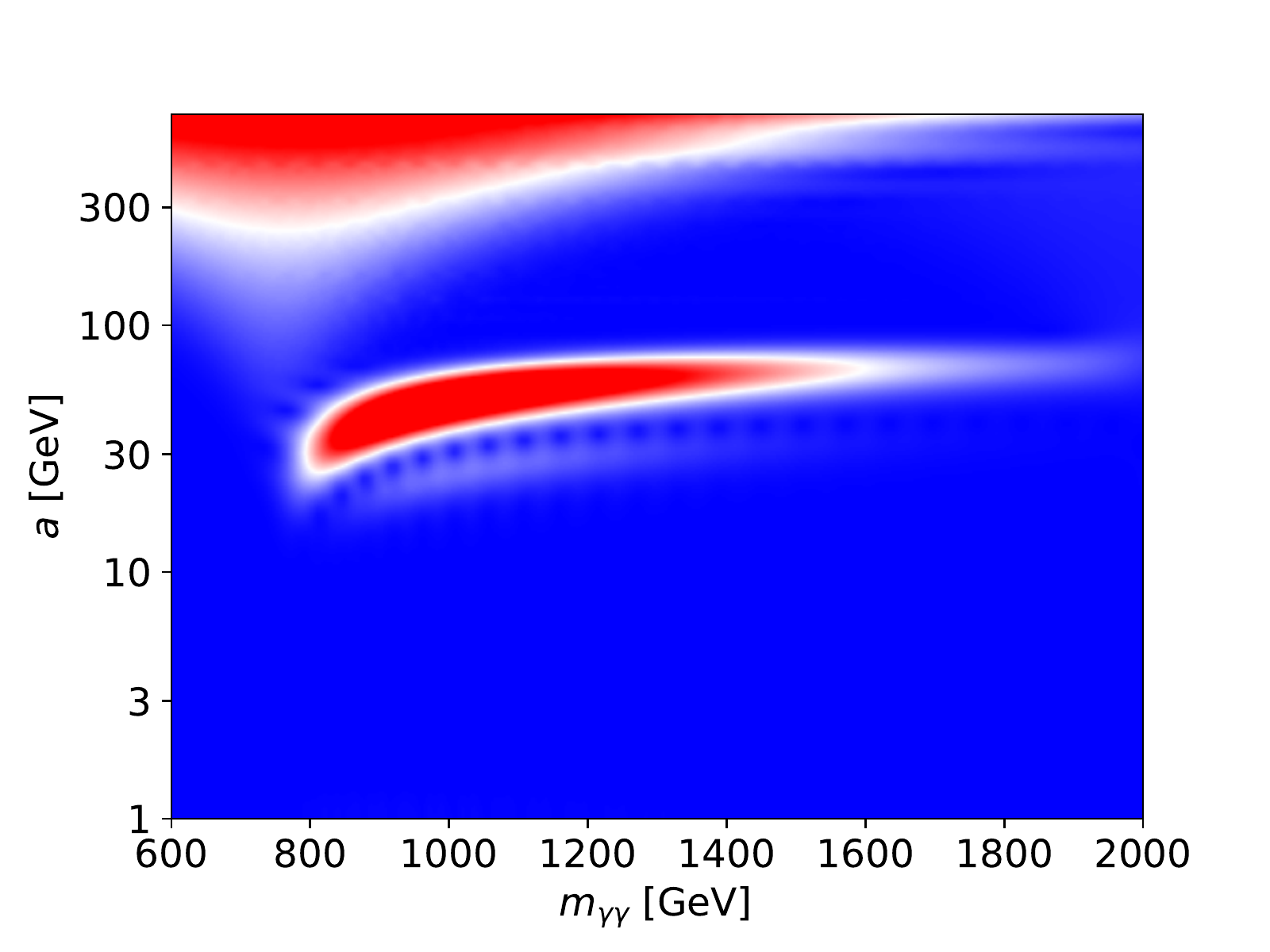}
    \caption{}
    \label{fig:SFix1}
  \end{subfigure}
  \begin{subfigure}{0.47\textwidth}
    \centering
    \includegraphics[width=\textwidth]{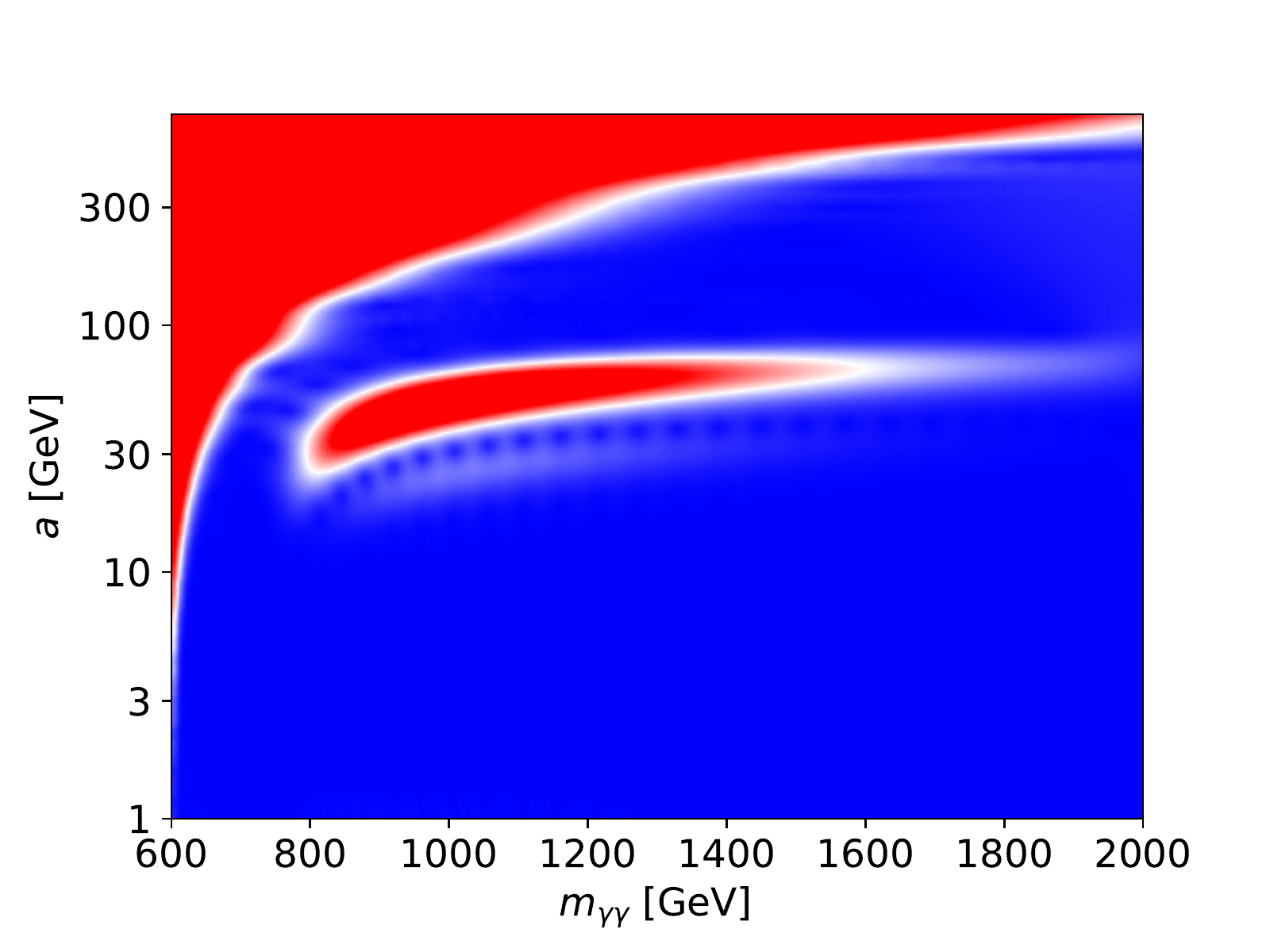}
    \caption{}
    \label{fig:SBFix1}
  \end{subfigure}
  ~
  \begin{subfigure}{0.47\textwidth}
    \centering
    \includegraphics[width=\textwidth]{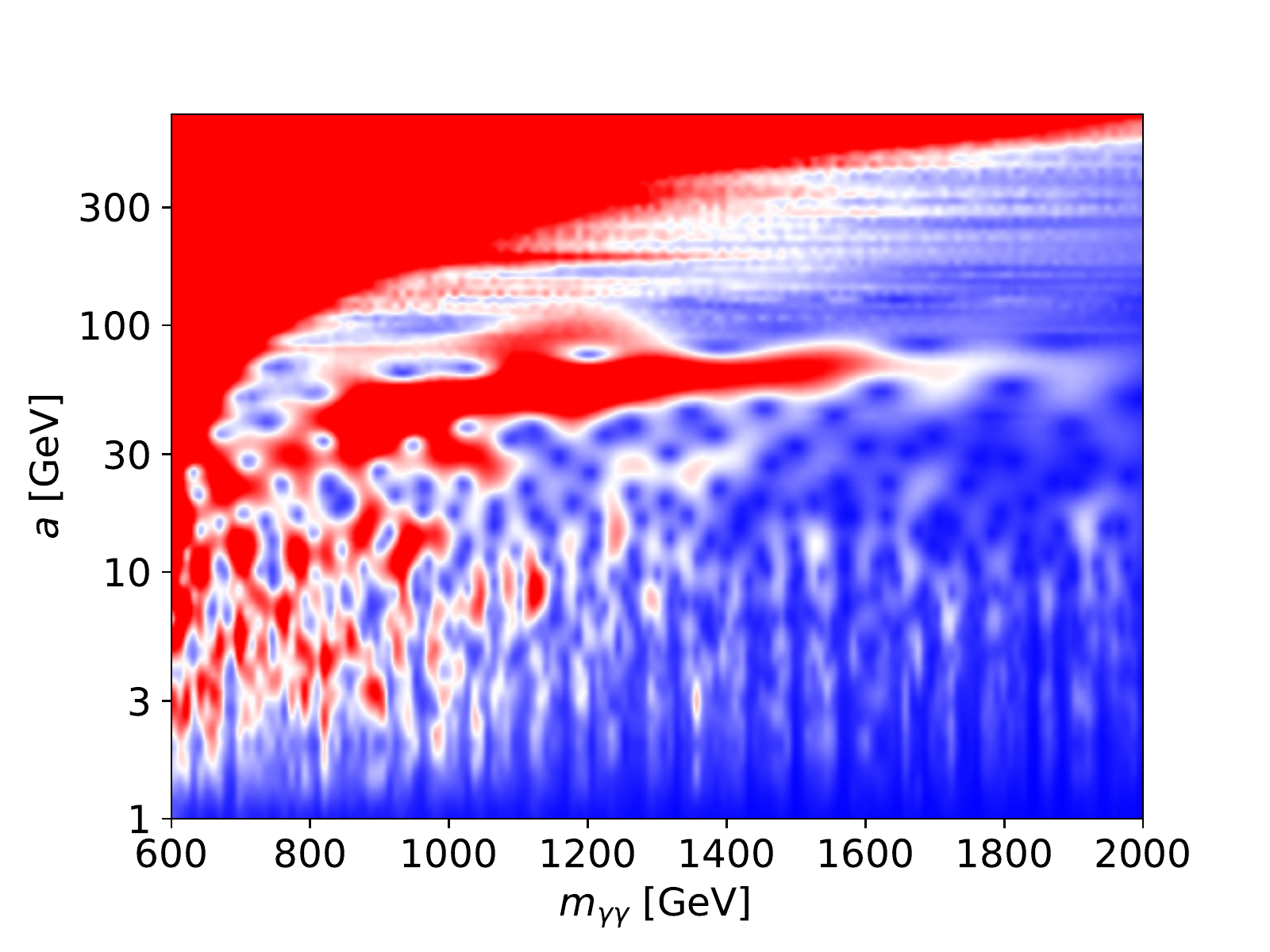}
    \caption{}
    \label{fig:RandomBS1}
  \end{subfigure}
  \begin{subfigure}{0.47\textwidth}
    \centering
    \includegraphics[width=\textwidth]{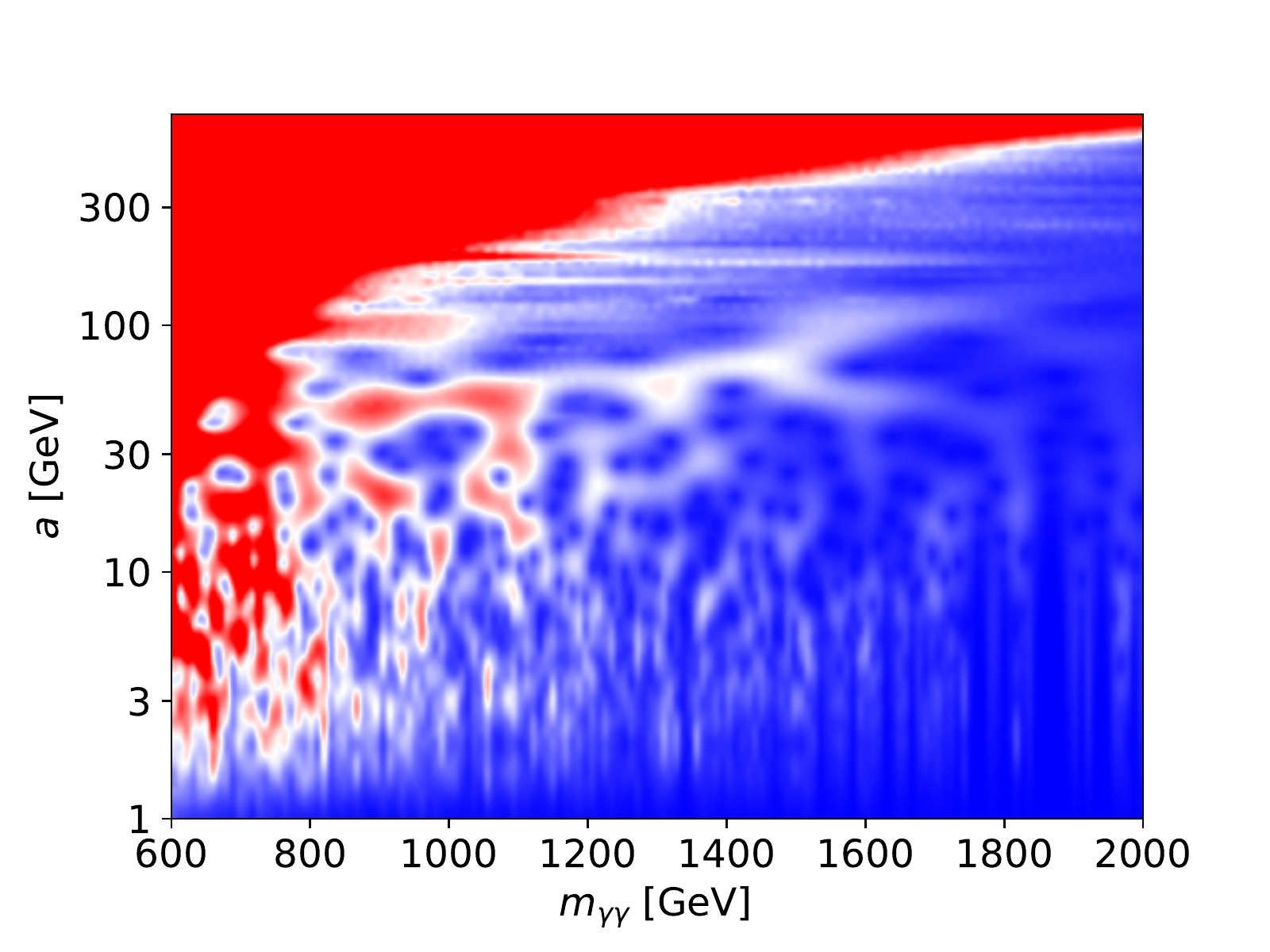}
    \caption{}
    \label{fig:RandomBS2}
  \end{subfigure}
  \captionsetup{justification=justified}
\caption{Examples of scalograms for the CW/LD with $k = 750$ GeV. (a) CWT of the smooth falling background from ref.~\cite{Aaboud:2017yyg} without statistical fluctuations. (b) CWT of the background with statistical fluctuations. (c) CWT of the signal alone, without statistical fluctuations, for $M_5 = 3$~TeV. (d) CWT of the same signal + background for $M_5 = 3$~TeV. (e) CWT of the signal + background with statistical fluctuations for $M_5 = 3$~TeV. (f)~CWT of the signal + background with statistical fluctuations for $M_5 = 5$~TeV.}\label{fig:ScalogramsEx}
\end{figure}

\begin{figure}[t]
  \begin{center}
    \includegraphics[width=0.85\textwidth]{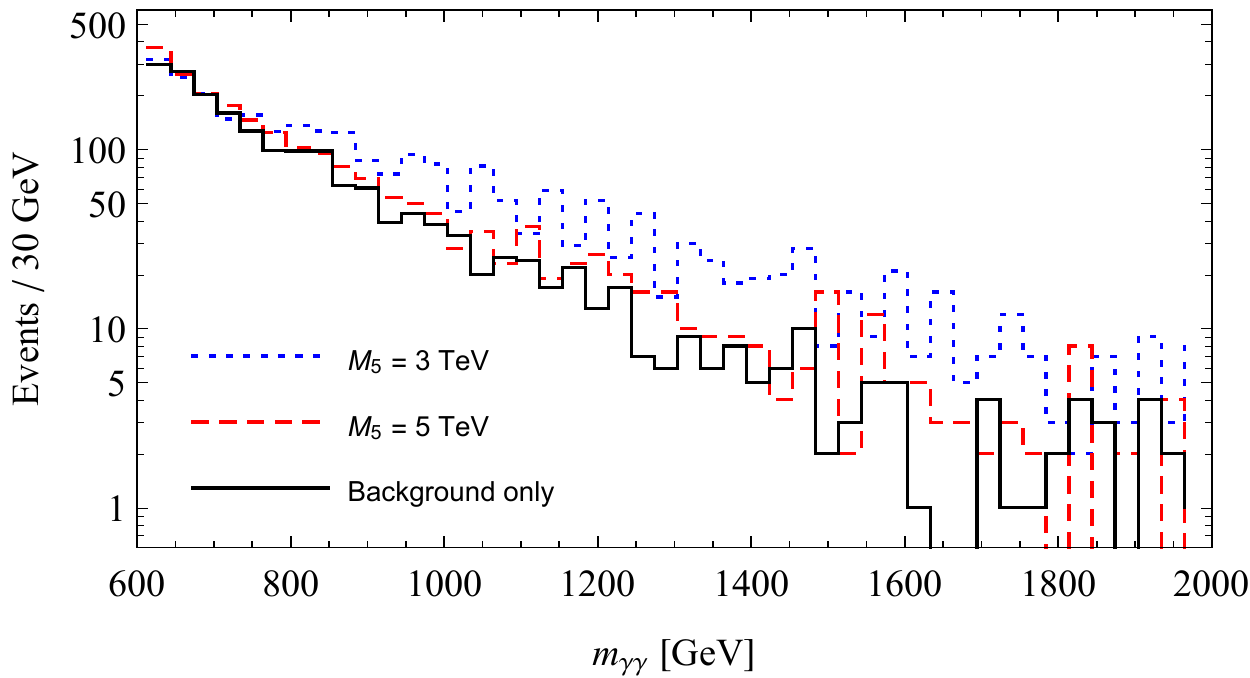}
  \end{center} 
  \caption{Mass spectra leading to the scalograms of figure~\ref{fig:ScalogramsEx}. The black (solid),  blue (dotted) and red (dashed) lines correspond to (b), (e) and (f), respectively.}\label{fig:ScalogramsExIn}
\end{figure}

Continuous Wavelet Transforms (CWT) address these issues by projecting a given signal over a basis of functions that are localized in both time and frequency space. An example of such a basis function is shown in the inset of figure~\ref{fig:Signal}. The output of the CWT is a scalogram, a two dimensional function which indicates how much a certain frequency is present at a given time. The CWT of the signal of figure~\ref{fig:Signal}, as well as a similar but weaker signal, is shown in figure~\ref{fig:ScalogramsEx} (with the input in figure~\ref{fig:ScalogramsExIn}). A signal that repeats itself with constant frequency would appear as a horizontal line, while one whose frequency changes with time as a line that moves up and/or down. A varying amplitude of the oscillations is also directly represented in the scalogram, as well as when the signal starts and ends. This makes the CWT very flexible when it comes to discovering generic periodic signals. 

While signals of the kind shown in figure~\ref{fig:Signal} can often be detected also directly by searches for resonances (e.g.,~\cite{Aaboud:2017yyg,Sirunyan:2018wnk,Aad:2019fac,ATLAS-CONF-2019-007,CMS-PAS-EXO-17-026}) or searches for continuum excesses at high masses (e.g.,~\cite{Aaboud:2017yyg,Sirunyan:2018wnk,Sirunyan:2018ipj}), the coverage of such searches is not robust. The sensitivity of searches for single resonant peaks on top of a smooth background is likely reduced by the presence of neighboring partly-overlapping peaks, in a model- and search-procedure-dependent fashion. Furthermore, the expected presence of multiple peaks is not being exploited in an optimal manner by such searches. Searches for continuum excesses will have reduced sensitivity for scenarios in which the signal does not extend to high masses or it makes comparable positive and negative contributions to the spectrum due to quantum interference (see, e.g.,~\cite{Djouadi:2016ack,Martin:2016bgw,Craig:2016iea}). In some cases, interference can even lead to pure deficits in the spectrum~\cite{Craig:2016iea}, challenging both the resonant and the continuum searches. It is also useful to note that the CWT analysis is almost insensitive to the systematic uncertainty on the normalization of the background mass spectrum. The CWT approach is thus complementary to the existing search strategies and is therefore worth exploring.

Although the amount of work on CWT in the context of collider searches has been limited up to now (see ref.~\cite{Spalla:2013abc}), they have been used in many different fields of science. For example, they have been used in astronomy~\cite{Lachowicz:2005abcd}, biology~\cite{Kiskin2018}, chaos theory~\cite{Benitez:2010abcd}, geophysics~\cite{Kumar:1997abcd, FLINCHEM2000177}, mechanical engineering~\cite{Heidary:2010abcd} and signal processing~\cite{1550194}. Work on the statistical significance of signals in CWT includes refs.~\cite{Lachowicz2009WaveletAA, npg-23-45-2016, npg-23-257-2016, PhysRevE.75.016707, Rouyer:2008abcd, angeo-25-2259-2007, Maraun:2004abcd, npg-22-139-2015, Torrence:1998abcd}.

The goal of the current paper is to demonstrate that continuous wavelet transforms can be used to discover periodic signals in kinematic distributions, in particular in the context of new physics searches at colliders. To do this, we will present a series of methods through which CWT can be used to discover such signals. These will range from the use of a simple test statistic to more advanced machine learning techniques. We find that CWT can compete with current methods and in some cases be sensitive to signals that would otherwise go undetected by current analyses.

The paper is organized as follows. We begin by defining the continuous wavelet transforms more carefully. The set of methods are then introduced. Windowed Fourier transforms are then presented to serve as a comparison. The different methods are then compared in the context of the diphoton signal of the CW/LD benchmark, details about which are provided in the appendix. Some concluding remarks complete the paper.

\section{Overview of continuous wavelet transforms}\label{Sec:CWT}
Assume $\psi(t)$ is a basis function localized in both time and frequency space. The continuous wavelet transform of a signal $f(t)$ at a scale $a > 0$ and translational parameter $b \in \mathbb{R}$ is given by a projection over rescaled and shifted version of $\psi(t)$:
\begin{equation}\label{eq:CWTdef}
  W(a, b) = \frac{1}{\sqrt{a}}\int_{-\infty}^{+\infty}f(t)\, \psi^*\left(\frac{t - b}{a}\right)dt \,.
\end{equation}
In practice, it is a measure of how much a certain frequency is present in the signal at a given time. The function $\psi(t)$ is known as the mother wavelet and its rescaled and shifted versions as daughter wavelets. The mother wavelet is required to satisfy two conditions:
\begin{equation}\label{eq:CWTCond1}
  \int_{-\infty}^{+\infty} |\psi(t)|^2 dt < \infty \,,
\end{equation}
\begin{equation}\label{eq:CWTCond2}
  c_\psi \equiv 2\pi \int_{-\infty}^{+\infty} \frac{|\Psi(\omega)|^2}{|\omega|}d\omega < \infty \,,
\end{equation}
where $\Psi(\omega)$ is the Fourier transform of $\psi(t)$. The first condition is simply that the mother wavelet has a finite norm and the second is known as the admissibility condition. The latter implies $\Psi(0)=0$, which means in turn that an admissible wavelet must integrate to zero and as such that the CWT of a constant function is zero. Note that for practical uses the signal might be binned, in which case the integral is replaced by a sum over bins.

We will use the \emph{Morlet wavelet} throughout this article. It consists of a localized wave packet and is given by:
\begin{equation}\label{eq:MorletDef}
  \psi(t) \equiv \frac{1}{\sqrt{B\pi}}\, e^{-t^2/B} \left(e^{i 2\pi C t} - e^{-\pi^2 B C^2}\right),
\end{equation}
where $B$ and $C$ are two constants that we will take as 2 and 1 respectively. With this choice, the wavelet transform of a signal will be maximum when its wavelength corresponds approximately to the scale $a$. The second term ensures that the admissibility condition is satisfied, though it can safely be ignored for our choice of parameters. The Morlet wavelet is shown in the inset of figure~\ref{fig:Signal}. Do note that there exist different conventions on the definition of the Morlet wavelet.\footnote{Our convention is chosen to conveniently match with that of the pywt Python package~\cite{pywt,pywt-publication}.}

In this article, we will be using the example of production of a set of resonances decaying to two photons. The invariant mass of the two photons $m_{\gamma\gamma}$ will play the role of $t$. When we discuss the distribution of $m_{\gamma\gamma}$ directly, we will say we are in mass space. When we deal with its wavelet transform, we will say we are in frequency space. The same ideas can be applied to dielectron, dimuon and other final states. 

\section{Search strategies with continuous wavelet transforms}\label{Sec:SearchStragetiesCWT}
We present in this section a series of methods through which continuous wavelet transforms can be used to discover periodic signals in kinematic distributions. These methods will be compared in section~\ref{Sec:Comparison} by applying them to the CW/LD scenario~\cite{Giudice:2017fmj} assuming the diphoton dataset of ref.~\cite{Aaboud:2017yyg} (37~fb$^{-1}$ at 13~TeV). All illustrations are also taken from examples of that model.

\subsection{Method 1: Model-specific search with a simple test statistic}\label{sSec:WTSS}
\begin{figure}[t!]
  \centering
  \begin{subfigure}{0.48\textwidth}
    \centering
    \includegraphics[width=\textwidth]{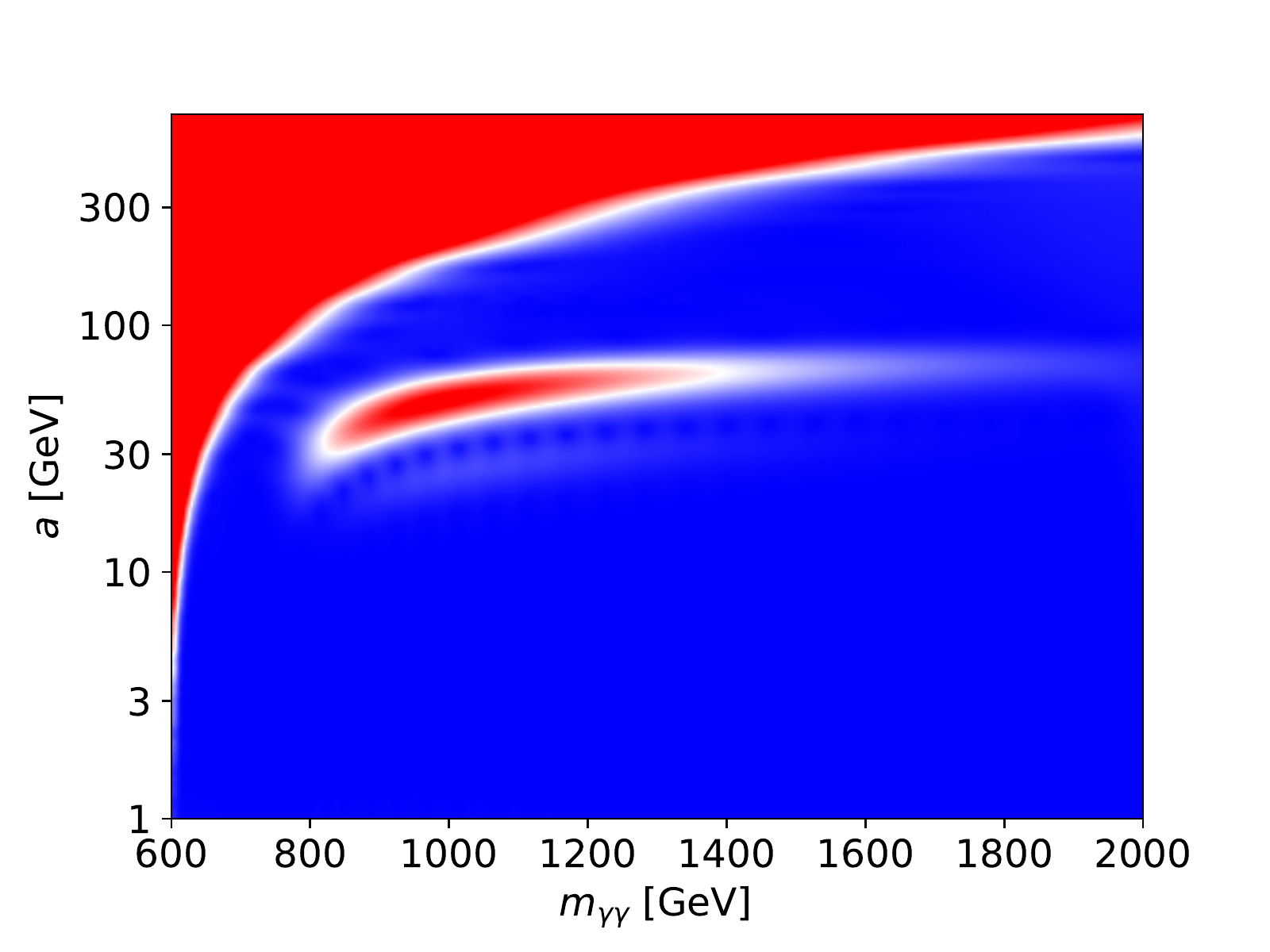}
    \caption{}
    \label{fig:pvalueMapP}
  \end{subfigure}
  ~
  \begin{subfigure}{0.48\textwidth}
    \centering
    \includegraphics[width=\textwidth]{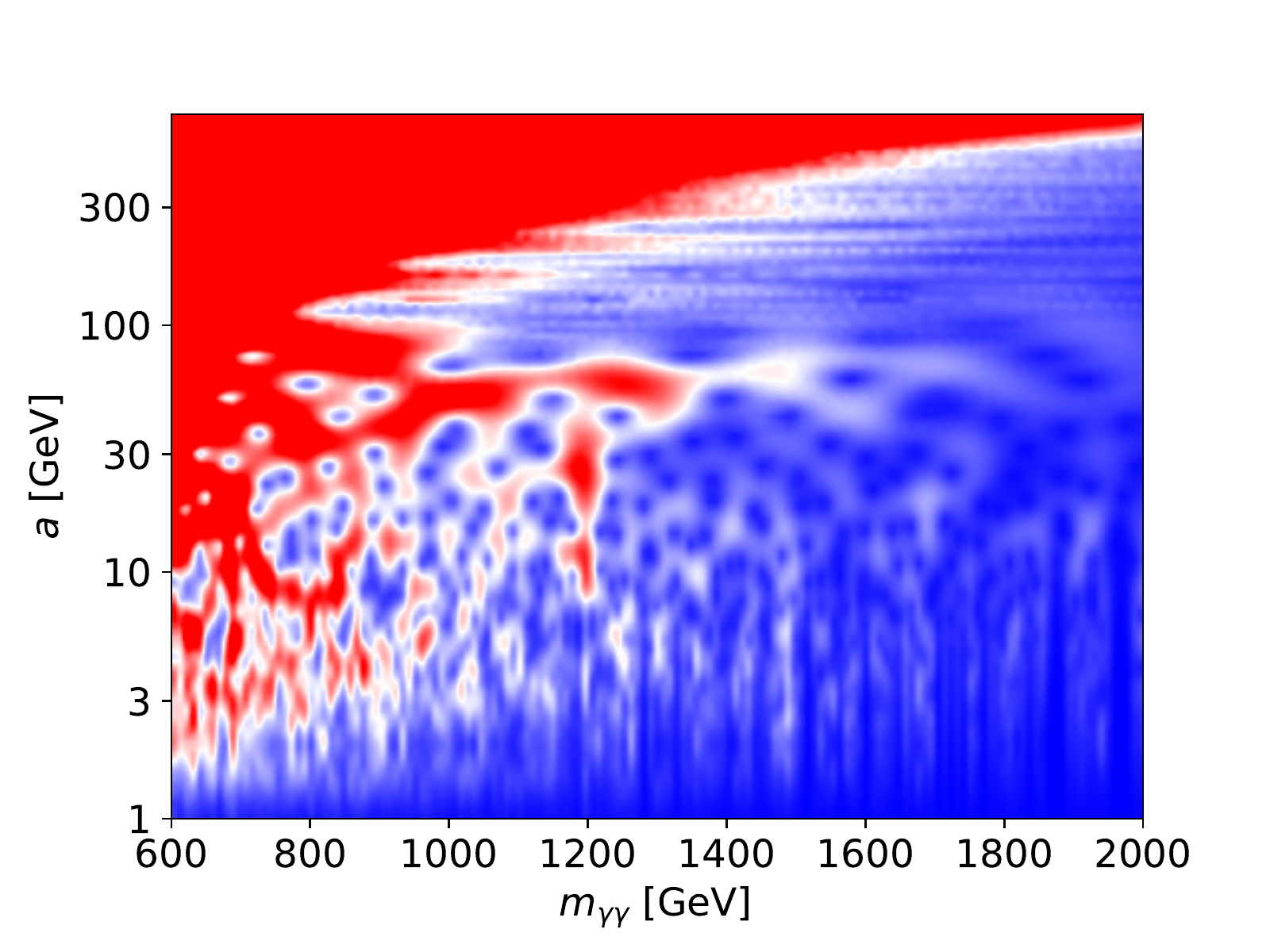}
    \caption{}
    \label{fig:pvalueMapIn}
  \end{subfigure}
  ~
    \begin{subfigure}{0.48\textwidth}
    \centering
    \includegraphics[width=\textwidth]{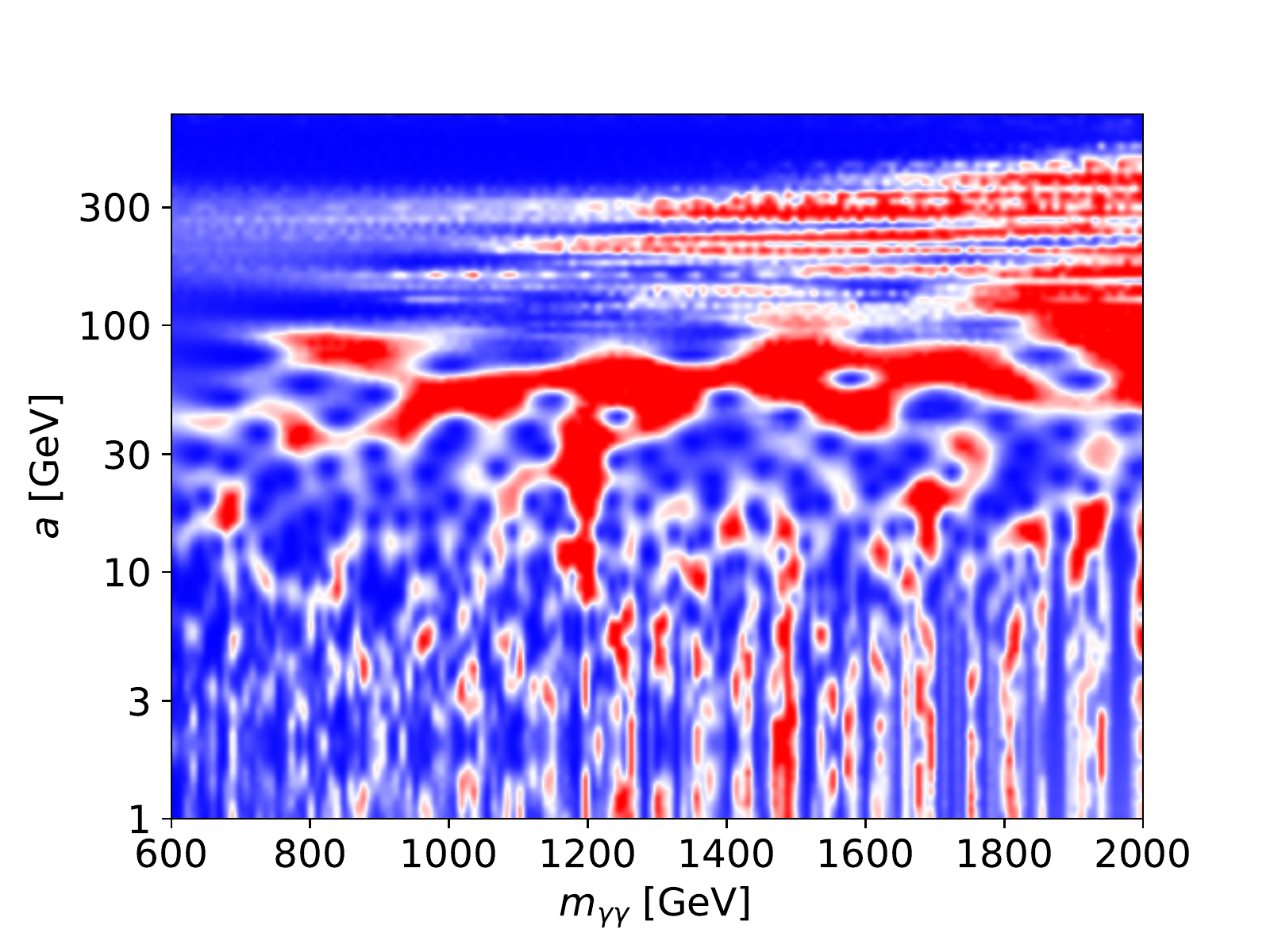}
    \caption{}
    \label{fig:pvalueMapOut}
  \end{subfigure}
  ~
    \begin{subfigure}{0.48\textwidth}
    \centering
    \includegraphics[width=\textwidth]{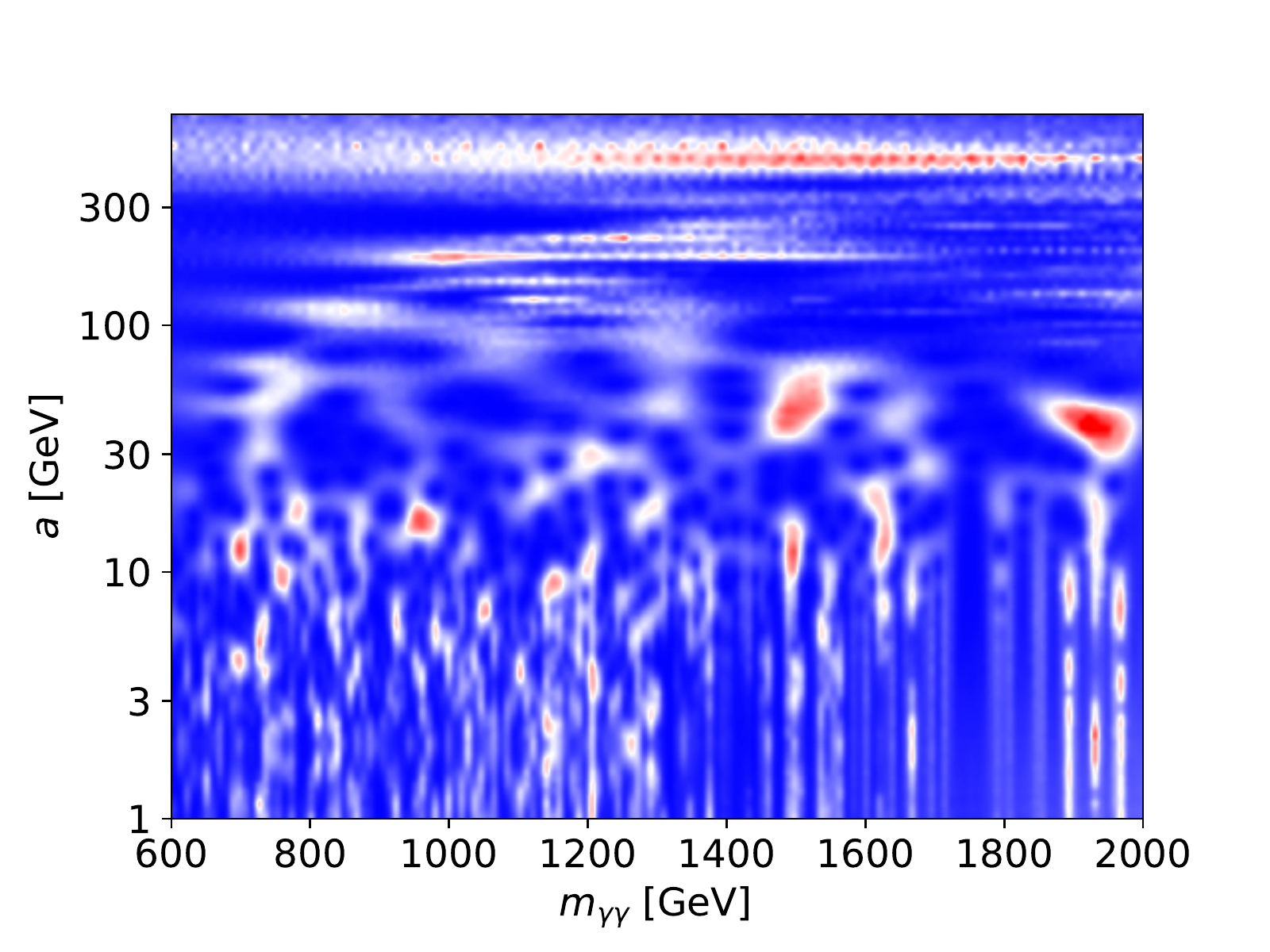}
    \caption{}
    \label{fig:pvalueMapOutB}
  \end{subfigure}
\caption{Example of (a) the CWT of the expected signal and background, (b) the same CWT with statistical fluctuations, (c) the corresponding $-\ln p$, and (d) $-\ln p$ of a typical background. The signal parameters used were $k = 750$~GeV and $M_5 = 4$~TeV.}\label{fig:pValueMap}
\end{figure}

The first strategy that we discuss is the use of a simple test statistic to detect a specific signal. A periodic signal is reflected in a scalogram as a series of ridges, each corresponding to a different harmonic. A given harmonic corresponds of course to a single scale for a given mass. Typically, the first harmonic is dominant and will have a large significance before the other harmonics are even visible. As such, we will concentrate on only the first one. Each point of a scalogram of the measured data can be assigned a local p-value by generating a large set of toy experiments with background only and determining which fraction of these  have a larger norm of the wavelet coefficient at that point. Statistical fluctuations are simulated by finely binning the expected spectrum in mass space, fluctuating the number of events in each bin according to the Poisson distribution, and then applying the CWT. Of course, the bin size must be chosen to be smaller than the expected scale of the signal. A signal will appear as a valley of low p-value, i.e.\ an extended structure. This can be seen in figure~\ref{fig:pValueMap}.
With these considerations in mind, a natural choice for a test statistic is:
\begin{equation}\label{eq:TestStatSS}
  t = -\sum_{i=i_{\text{min}}}^{i_{\text{max}}}\frac{1}{a_i}\ln p_i(a_i) \,,
\end{equation}
where the sum is over the mass bins of the scalogram, $a_i$ is the scale of the first harmonic for bin $i$ and $p_i(a_i)$ the local p-value for bin $i$ at scale $a_i$. Roughly speaking, $i_{\text{min}}$ is where the first harmonic starts to be discernible and $i_{\text{max}}$ where it stops to be. In practice, it is best to perform an optimization of these parameters on a case by case basis. The division by the scale is performed to counteract the fact that fluctuations typically span a mass range of the order of their scale. This test statistic is also inspired by the Fisher method \cite{Fischer}. More advanced test statistics could in principle be used, but eq.~(\ref{eq:TestStatSS}) is easy to implement and will be shown to return good limits.

\subsection{Method 2: Model-independent search with a simple test statistic}\label{sSec:GenericSignal}
Arguably the greatest strength of continuous wavelet transforms is that they can reveal a periodic signal that was not predicted by any previously considered model. If an anomalous region is present in a scalogram, a question that would need to be answered is how significant it is. Previous work on the subject includes~\cite{npg-22-139-2015, npg-23-45-2016}.

First, bins of interest can be selected by asking that their local p-value be below a certain value. They are then grouped into continuous regions. The test statistic (\ref{eq:TestStatSS}) is then applied to each region by taking the bin with the smallest p-value of each column. The largest test statistic is kept as a hyperstatistic. The statistical significance can then be obtained via a series of toy experiments.

\subsection{Method 3: Neural network and a simple test statistic}\label{sSec:RegionFinder}

Wavelet transforms map a periodic signal present over a background to an excess over an extended region in a scalogram. When the amount of statistics available is limited, such excesses can potentially be mistaken for simple statistical fluctuations of the background. As can be seen in figure~\ref{fig:ScalogramsEx}, the scalogram of a given signal can take a very complicated form in practice. However, it is clear that a real signal will tend to present certain features that are typically absent from background fluctuations and vice-versa. These features can potentially be used to increase the statistical significance of a given signal. Trying to manually classify them is at the very least an extremely daunting task, but it is an obvious application of machine learning.

\begin{table}[t]
	\begin{subfigure}{.5\linewidth}
		{\footnotesize
			\setlength\tabcolsep{5pt}
			\begin{center}
				\begin{tabular}{ll}
					\toprule
					Layer                  & Parameters \\
					\cmrule
					Input layer            & 63 mass bins $\times$ 56 scale bins \\
					Convolutional layer 1  & $\#$ filters = 32                   \\
					& kernel size = (5, 5)                \\
					& Activation: Softplus                \\
					MaxPooling 1           & Pooling size = (2, 2)               \\
					Convolutional layer 2  & $\#$ filters = 64                   \\
					& kernel size = (5, 5)                \\
					& Activation: Softplus                \\
					MaxPooling 2           & Pooling size = (2, 2)               \\
					Convolutional layer 3  & $\#$ filters = 128                  \\
					& kernel size = (5, 5)                \\
					& Activation: Softplus                \\
					Dense 1                & $\#$ of nodes = 5000                \\
					& Activation: Softplus                \\
					Output layer           & $\#$ of nodes = 3528                \\
					\bottomrule
				\end{tabular}
			\end{center}
		}
		\caption{}
	\end{subfigure}
	\begin{subfigure}{.5\linewidth}
		{\footnotesize
			\setlength\tabcolsep{5pt}
			\begin{center}
				\begin{tabular}{ll}
					\toprule
					Setting                   & Choice \\
					\cmrule
					Optimizer                 & \textsc{Adam}        \\
					Loss function             & Mean squared error   \\
					$\#$ training experiments & 5000                 \\
					Validation split          & 0.2                  \\
					Batch size                & 1000                 \\
					$\#$ epochs               & 200                  \\
					Callback                  & Smallest validation  \\
					& loss function        \\
					\bottomrule
				\end{tabular}
			\end{center}
		}
		\caption{}
	\end{subfigure}
	\caption{(a) Structure of the convolutional neural network for the region finder. (b) Training parameters. All parameters not specified in these tables are left at their default \textsc{Keras} values.} 
	\label{table:NNstructureRF}
\end{table}

\begin{figure}[t!]
	\centering
	\begin{subfigure}{0.48\textwidth}
		\centering
		\includegraphics[width=\textwidth]{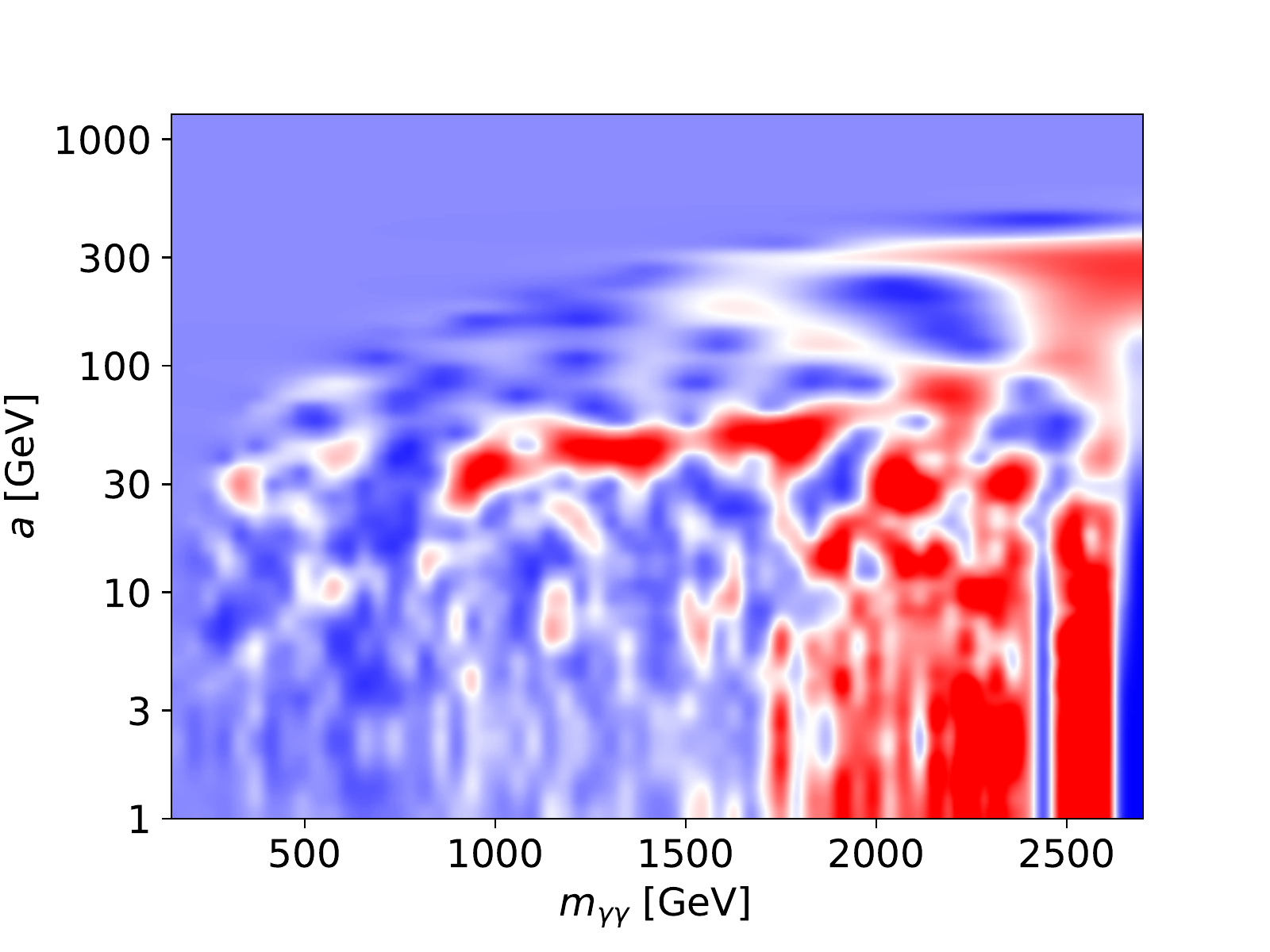}
		\caption{}
		\label{fig:TestIn}
	\end{subfigure}
	~
	\begin{subfigure}{0.48\textwidth}
		\centering
		\includegraphics[width=\textwidth]{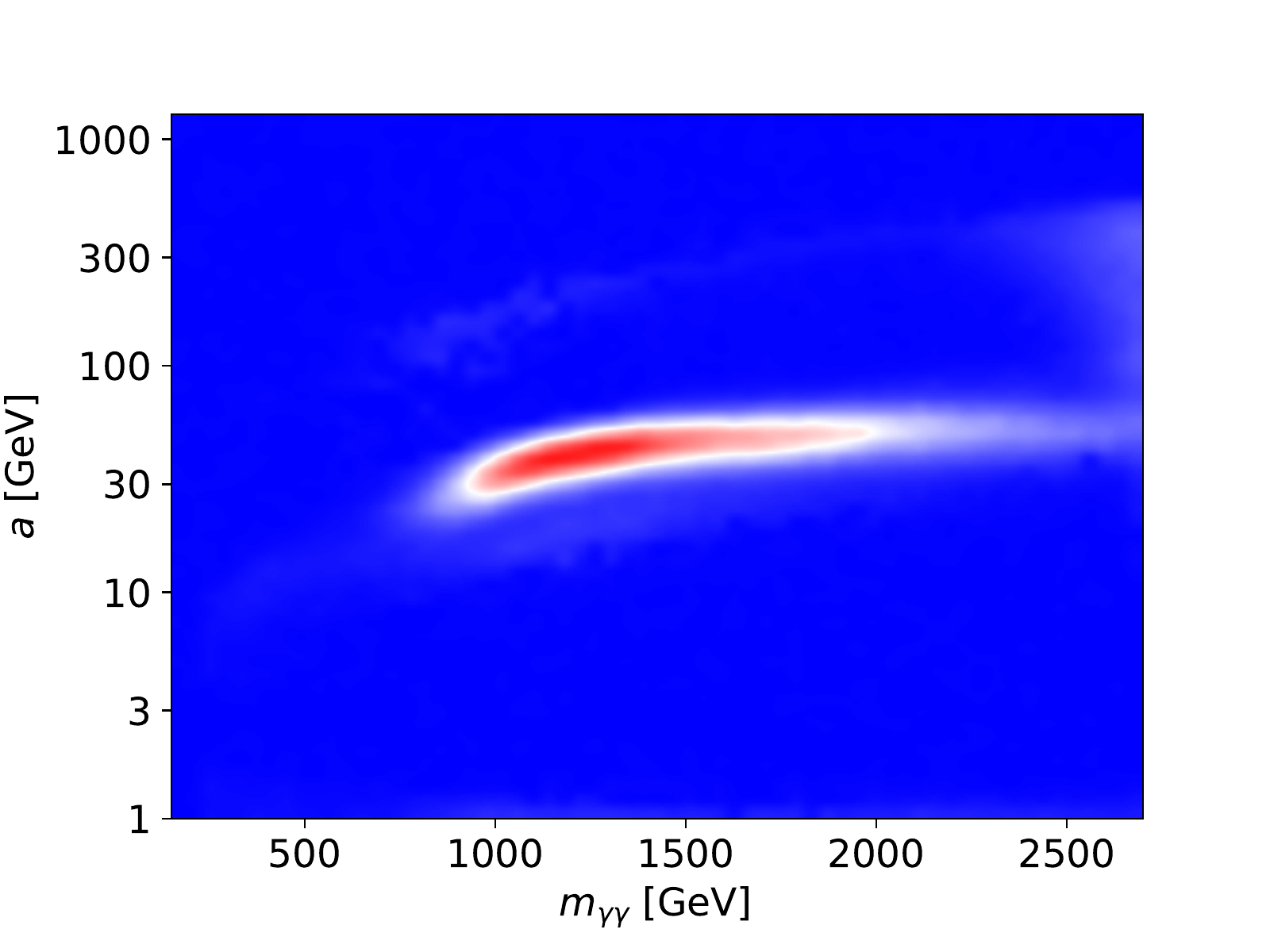}
		\caption{}
		\label{fig:TestOut}
	\end{subfigure}
	\caption{Example of the region finder algorithm input and output. (a) The input, which is the norm of the CWT divided by the expectation value of the background. (b) The output, which is trained to be $|W_{s,\text{exp}}|/\langle |W_b| \rangle$. The parameters used were $k = 750$~GeV and $M_5 = 4$~TeV. The original signal is the same as in figure~\ref{fig:pvalueMapP}.}\label{fig:TestInOut}
\end{figure}

One possibility is to train a neural network to identify regions compatible with the signal searched for inside a scalogram and then calculate their significance using a test statistic. Note that after the mass spectrum is sampled with sufficient resolution to capture the details of the signal, the resulting scalogram can be sampled in a cruder fashion thanks to the extended nature of the excess. This helps making the size of the neural network's input manageable. The neural network is trained on an equal mix of pure backgrounds and examples with signals from random points in the parameter space of the model (with signal strength above some optimized threshold). In both cases, the network's input is the norm of the CWT of the signal + background divided by the expectation value of the norm of the CWT of the background only $\langle |W_b| \rangle$. When there is a signal, the output it is trained to return is $|W_{s,\text{exp}}|/\langle |W_b| \rangle$, where $|W_{s,\text{exp}}|$ is the norm of the wavelet transform of the expected signal. When there is no signal, the neural network is trained to return zero in every bin. The details of the neural network we used are given in table~\ref{table:NNstructureRF}. When applied to a pseudo-experiment, the neural network will assign bins compatible with a signal-like excess a much larger value than those associated to the background. An example of this for a sample containing a signal is shown in figure~\ref{fig:TestInOut}, which shows that the neural network returns a very signal-like shape. For background-only samples, the neural network typically fails to return such well-defined regions. In principle, one could also train the neural network to return some other function, as long as its region of maximal value corresponds to where the excess should appear in the scalogram. All neural networks were implemented via the \textsc{Python} deep learning library \textsc{Keras}~\cite{chollet2015keras} with the \textsc{TensorFlow} backend~\cite{Abadi:2016kic} using the \textsc{Adam} optimizer~\cite{Kingma:2014vow}.

Having identified the region of interest, one can select for each mass column the scale with the largest neural network output and apply the test statistic of eq.~(\ref{eq:TestStatSS}) using the actual value of the wavelet coefficient for this bin. A minimal value of the neural network output is required for a bin to be counted in the test statistic. This way, the neural network's assessment of whether a signal-like excess is present at all, is taken into account. We scan the value of this threshold to maximize significance. The expected sensitivity limits are obtained by pseudo-experiments.

\subsection{Method 4: Classifier neural network}\label{sSec:Classifier}

A more powerful way to use machine learning to discover a specific signal in a scalogram is with a classifier. First, toy experiments, with and without the signal (for a specific choice of model parameters), are generated and finely binned mass spectra are produced. The continuous wavelet transform of the mass spectrum is taken and rebinning is performed to make the size of the neural network's input manageable. The norm of each bin is then passed as an input to a convolutional neural network whose output is trained to be one when the signal is present and zero otherwise. The output of the neural network can then be used as a test statistic. The structure of the neural network that we used and additional parameters are provided in table~\ref{table:NNstructureClassifier}.

\begin{table}[t]
\begin{subfigure}{.5\linewidth}
{\footnotesize
\setlength\tabcolsep{5pt}
\begin{center}
\begin{tabular}{ll}
\toprule
Layer                  & Parameters \\
\cmrule
Input layer            & 63 mass bins $\times$ 56 scale bins \\
Convolutional layer 1  & $\#$ filters = 4                    \\
                       & kernel size = (3, 3)                \\
                       & Activation: Elu                     \\
MaxPooling 1           & Pooling size = (2, 2)               \\
Convolutional layer 2  & $\#$ filters = 8                    \\
                       & kernel size = (3, 3)                \\
                       & Activation: Sigmoid                 \\
MaxPooling 2           & Pooling size = (2, 2)               \\
Convolutional layer 3  & $\#$ filters = 16                   \\
                       & kernel size = (3, 3)                \\
                       & Activation: Sigmoid                 \\
Dense 1                & $\#$ of nodes = 200                 \\
                       & Activation: Sigmoid                 \\
Dense 2                & $\#$ of nodes = 100                 \\
                       & Activation: Sigmoid                 \\
Output layer           & $\#$ of nodes = 1                   \\
\bottomrule
\end{tabular}
\end{center}
}
\caption{}
\end{subfigure}
\begin{subfigure}{.5\linewidth}
{\footnotesize
\setlength\tabcolsep{5pt}
\begin{center}
\begin{tabular}{ll}
\toprule
Setting                   & Choice \\
\cmrule
Optimizer                 & \textsc{Adam}        \\
Loss function             & Binary cross entropy \\
$\#$ training experiments & 4000                 \\
Validation split          & 0.2                  \\
Batch size                & 1000                 \\
$\#$ epochs               & 500                  \\
Callback                  & Smallest validation  \\
                          & loss function        \\
\bottomrule
\end{tabular}
\end{center}
}
\caption{}
\end{subfigure}
\caption{(a) Structure of the convolutional neural network of the wavelet classifier. (b) Training parameters. All parameters not specified in these tables are left at their default \textsc{Keras} values.} 
\label{table:NNstructureClassifier}
\end{table}

\subsection{Method 5: Autoencoder neural network}\label{sSec:Autoencoders}
\begin{figure}[t!]
  \centering
   \captionsetup{justification=centering}
  \begin{subfigure}{0.48\textwidth}
    \centering
    \includegraphics[width=\textwidth]{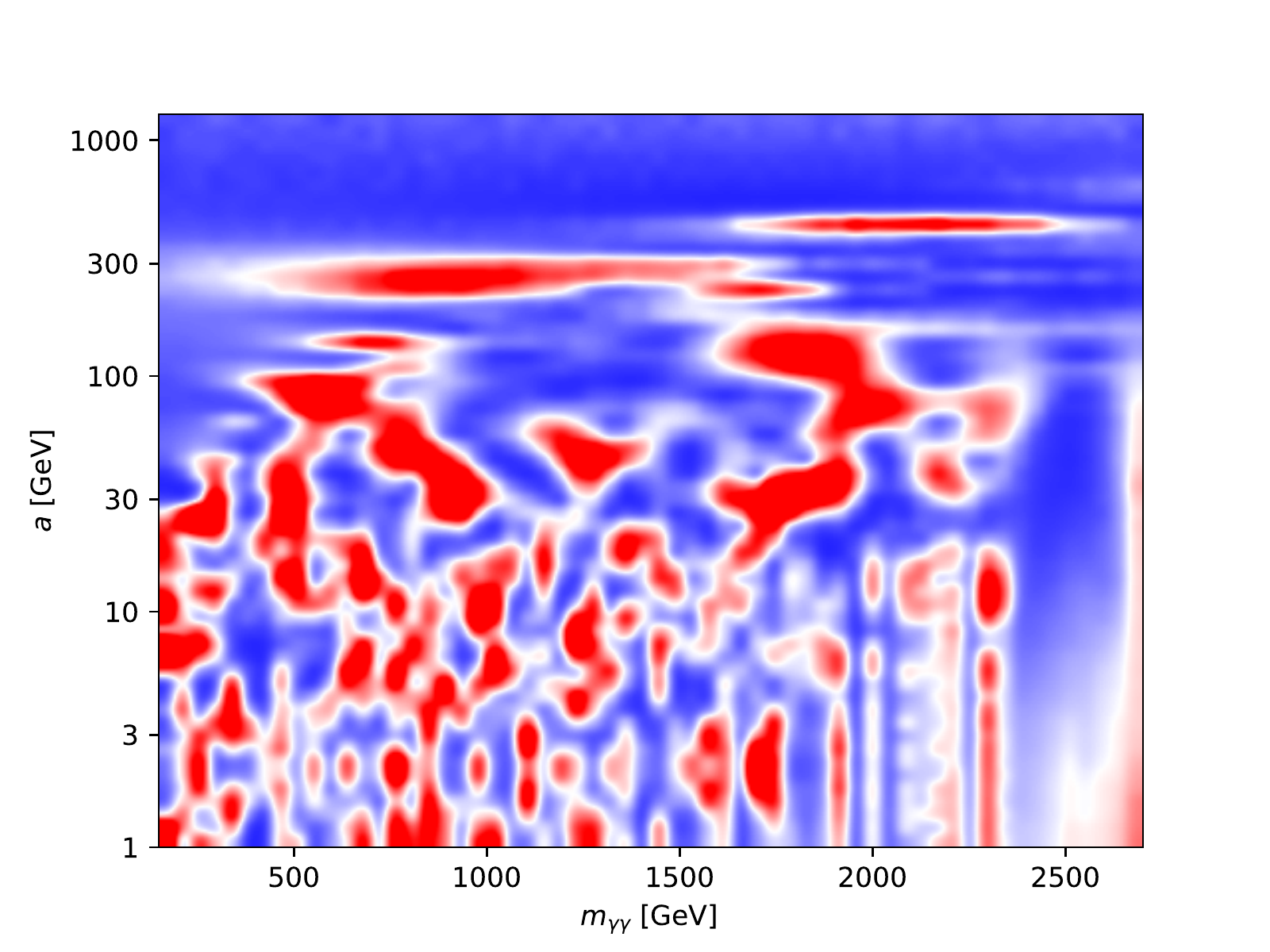}
    \caption{}
    \label{fig:SAENoSIn}
  \end{subfigure}
  ~
  \begin{subfigure}{0.48\textwidth}
    \centering
    \includegraphics[width=\textwidth]{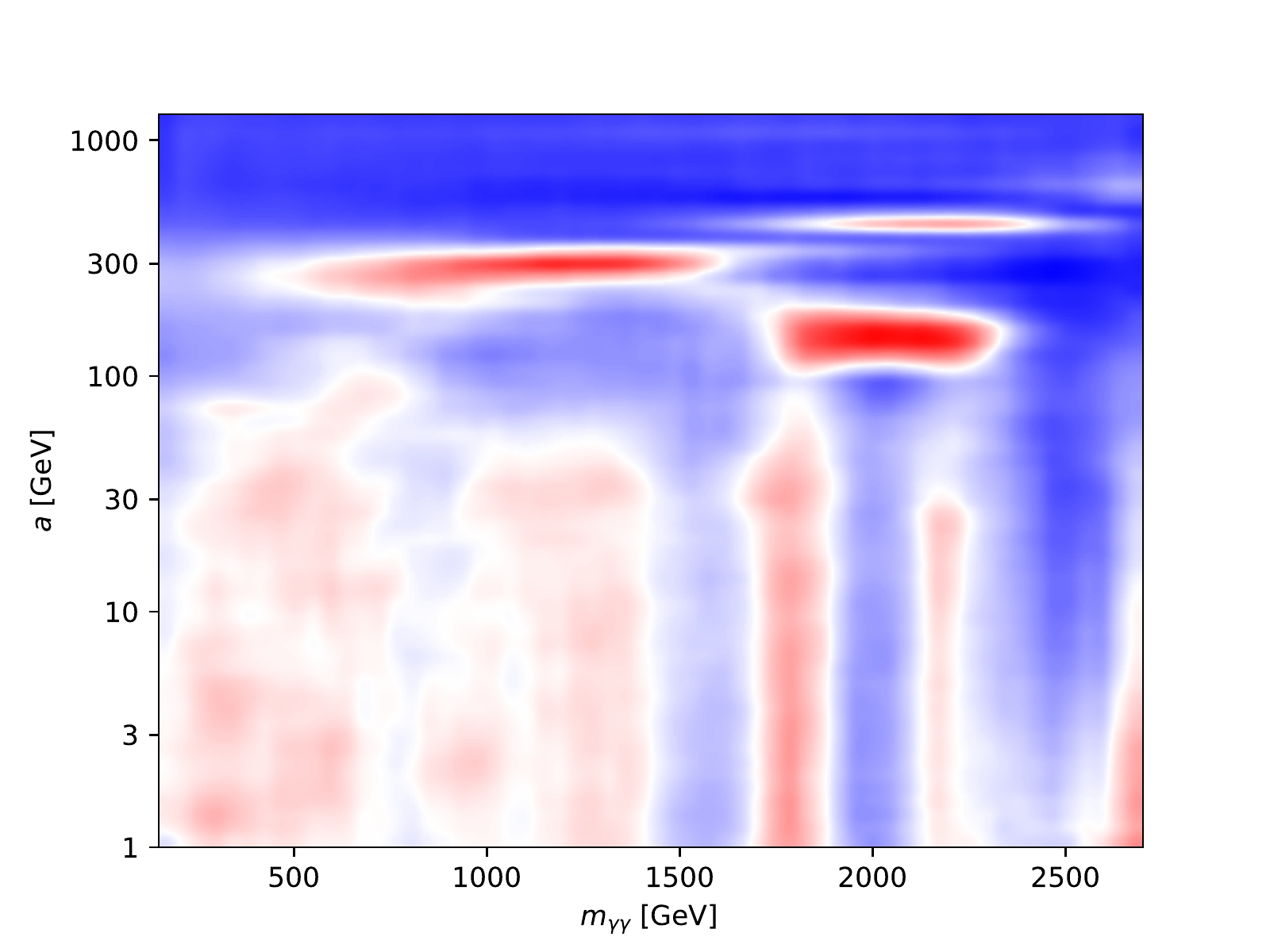}
    \caption{}
    \label{fig:SAENoSOut}
  \end{subfigure}
  ~
  \begin{subfigure}{0.48\textwidth}
    \centering
    \includegraphics[width=\textwidth]{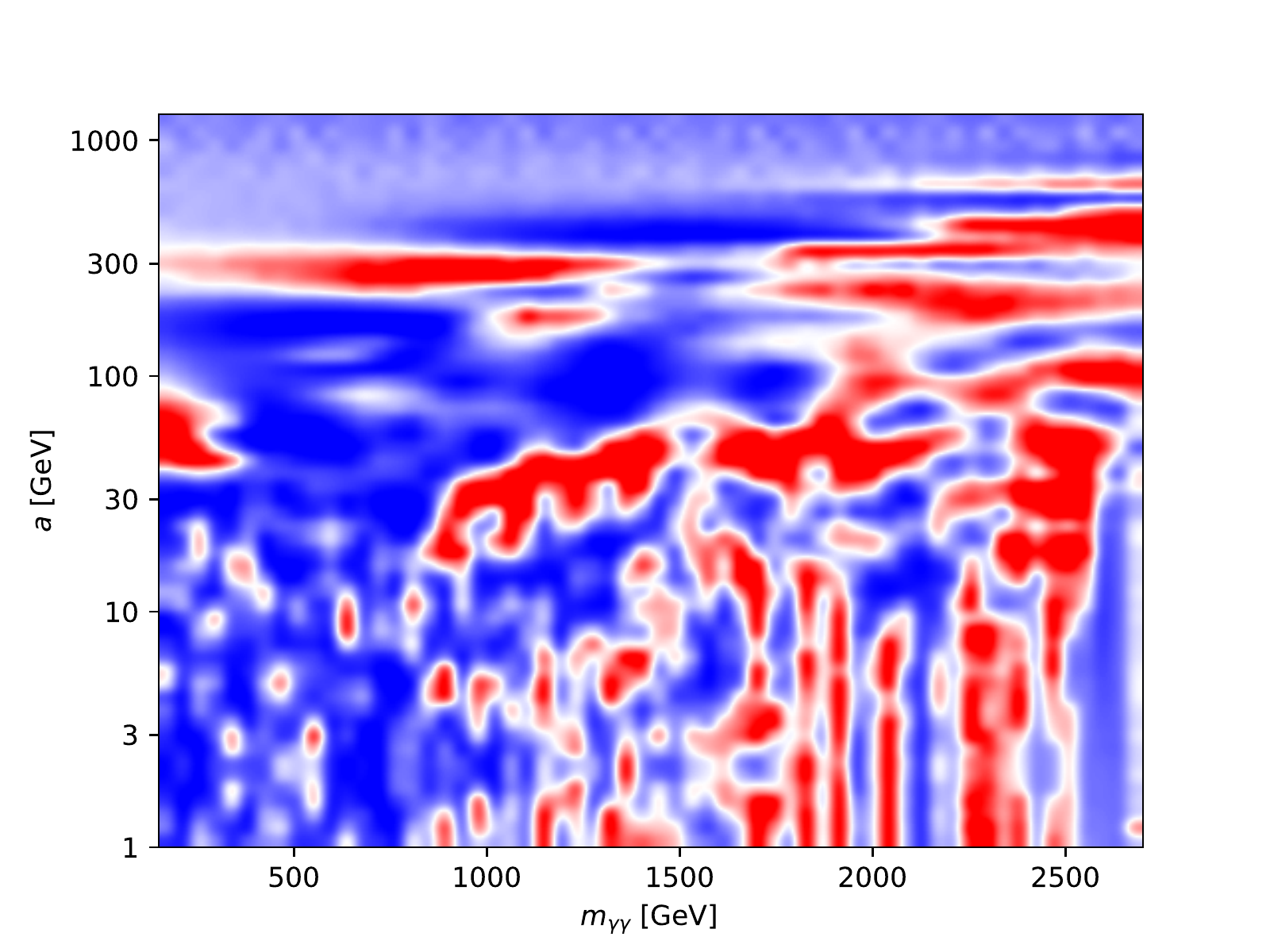}
    \caption{}
    \label{fig:SAESIn}
  \end{subfigure}
  \begin{subfigure}{0.48\textwidth}
    \centering
    \includegraphics[width=\textwidth]{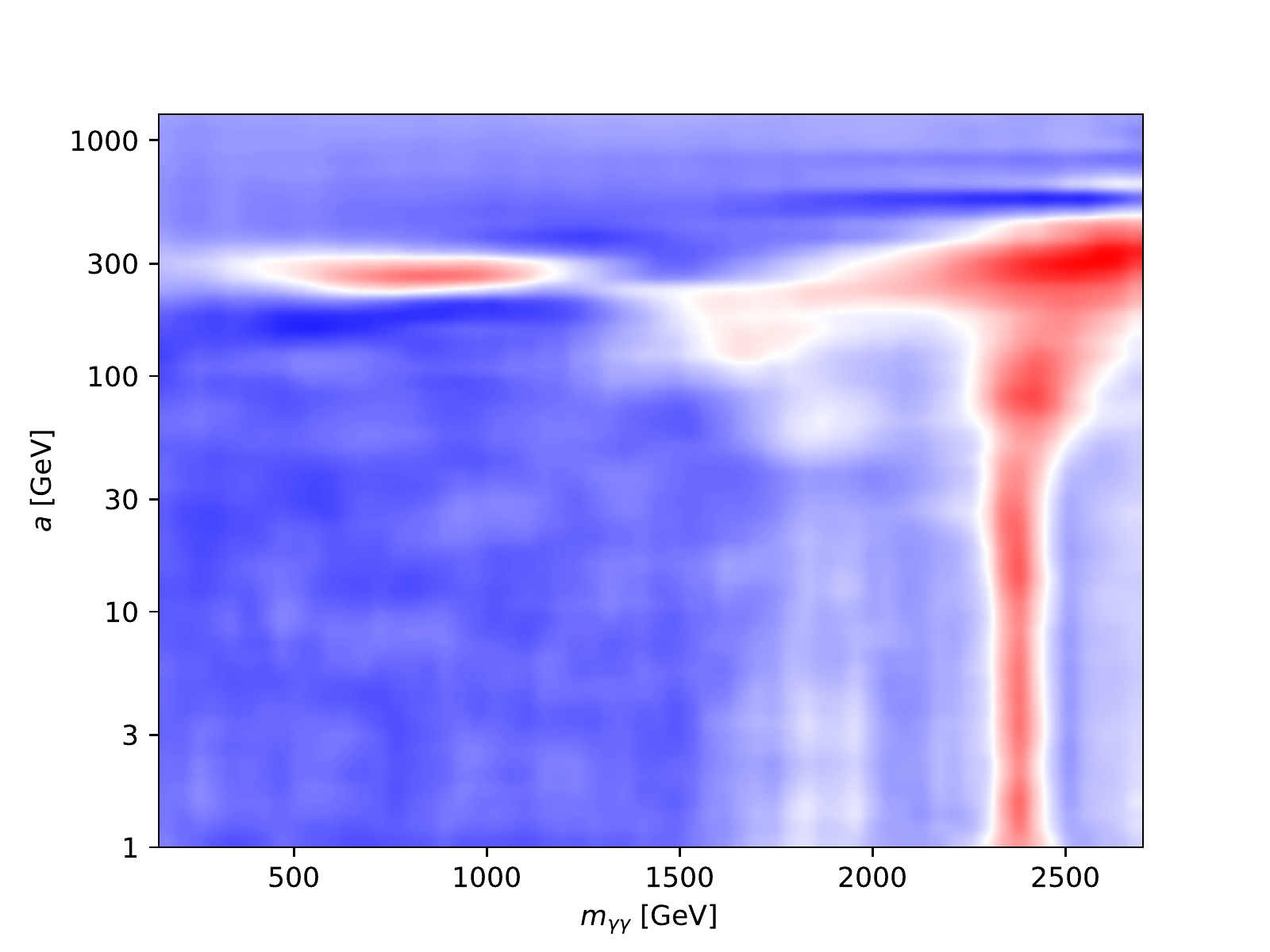}
    \caption{}
    \label{fig:SAESOut}
  \end{subfigure}
  \captionsetup{justification=justified}
\caption{Examples of the input and output of the autoencoder. (a)$\to$(b): Background-only sample and its reconstruction after encoding. (c)$\to$(d): Sample containing a signal and its (mis)reconstruction after encoding. The parameters used for the signal were $M_5=4$~TeV and $k=750$ GeV. The original signal is shown figure~\ref{fig:pvalueMapP}.}\label{fig:ScalogramsAutoencoder}
\end{figure}

A more model-independent option to look for anomalies in scalograms is via autoencoders, similar to their application to jet images in refs.~\cite{Heimel:2018mkt,Farina:2018fyg}. The idea is to use an autoencoder network to compress a scalogram to a smaller set of parameters which are then used to reconstruct the original scalogram. The neural network is then trained on backgrounds only to reproduce the original scalogram as well as possible. After training, the neural network should be able to reproduce the original scalogram to good approximation if applied to a typical background sample, and fail if applied to a sample that contains a signal. One can then use the reconstruction loss function as a test statistic. The details of the neural network are presented in table~\ref{table:NNstructureAE} and are simply an adaption of the convolutional neural network of ref.~\cite{Farina:2018fyg}. The input of the neural network, which is also the output it is trained to return, is the negative log of the local p-value of each bin. Examples of inputs and outputs of the neural network are shown in figure~\ref{fig:ScalogramsAutoencoder}. As can be seen, the autoencoder manages to reproduce approximately the major fluctuations of the background. At the same time, as desired, it fails to reproduce the signal, which results in a much larger value for the reconstruction loss function.

\begin{table}[t]
\begin{subfigure}{.5\linewidth}
{\footnotesize
\setlength\tabcolsep{5pt}
\begin{center}
\begin{tabular}{ll}
\toprule
Layer                    & Parameters \\
\cmrule
Input layer              & 60 mass bins $\times$ 56 scale bins \\
Convolutional layer 1    & $\#$ filters = 128                  \\
                         & kernel size = (3, 3)                \\
                         & Activation: Elu                     \\
MaxPooling 1             & Pooling size = (2, 2)               \\
Convolutional layer 2    & $\#$ filters = 128                  \\
                         & kernel size = (3, 3)                \\
                         & Activation: Elu                     \\
MaxPooling 2             & Pooling size = (2, 2)               \\
Convolutional layer 3    & $\#$ filters = 128                  \\
                         & kernel size = (3, 3)                \\
                         & Activation: Elu                     \\
Dense 1                  & $\#$ of nodes = 40                  \\
                         & Activation: Elu                     \\
Dense 2 (\emph{Encoded}) & $\#$ of nodes = 20                  \\
                         & Activation: Elu                     \\
Dense 3                  & $\#$ of nodes = 40                  \\
                         & Activation: Elu                     \\
Convolutional layer 4    & $\#$ filters = 128                  \\
                         & kernel size = (3, 3)                \\
                         & Activation: Elu                     \\
UpSampling 1             & Upsampling factors = (2, 2)         \\
Convolutional layer 5    & $\#$ filters = 128                  \\
                         & kernel size = (3, 3)                \\
                         & Activation: Elu                     \\
UpSampling 2             & Upsampling factors = (2, 2)         \\
Convolutional layer 6    & $\#$ filters = 1                    \\
                         & kernel size = (3, 3)                \\
                         & Activation: Elu                     \\
Output layer             & 60 mass bins $\times$ 56 scale bins \\
\bottomrule
\end{tabular}
\end{center}
}
\caption{}
\end{subfigure}
\begin{subfigure}{.5\linewidth}
{\footnotesize
\setlength\tabcolsep{5pt}
\begin{center}
\begin{tabular}{ll}
\toprule
Setting                   & Choice \\
\cmrule
Optimizer                 & \textsc{Adam}        \\
Loss function             & Mean squared error   \\
$\#$ training experiments & 5000                 \\
Validation split          & 0.2                  \\
Batch size                & 1000                 \\
Padding                   & Same                 \\
$\#$ epochs               & 100                  \\
Callback                  & Smallest validation  \\
                          & loss function        \\
\bottomrule
\end{tabular}
\end{center}
}
\caption{}
\end{subfigure}
\caption{(a) Structure of the convolutional neural network for the autoencoder. (b) Training parameters. All parameters not specified in these tables are left at their default \textsc{Keras} values.} 
\label{table:NNstructureAE}
\end{table}

\subsection{Fourier analysis as a reference}\label{Sec:FourierAnalysis}
Before moving on to comparing the methods, we discuss the use of Fourier transforms to discover periodic signals, which is the approach that was proposed in ref.~\cite{Giudice:2017fmj} in the context of the CW/LD scenario. This will serve as a reference, though it is clear that CWT are far more general.

We define in general the power spectrum as:
\begin{equation}\label{eq:PowerSpectrum}
  P(T) = \left|\frac{1}{\sqrt{2\pi}}\int_{m_{\text{min}}}^{m_{\text{max}}}dm\frac{d\sigma}{dm}\frac{1}{L}\exp\left(i\frac{2\pi g(m)}{T}\right)\right|^2.
\end{equation}
The quantity $L$ is the parton luminosity given for CW/LD by $L(m^2) = \mathcal{L}_{gg}(m^2) + \frac{4}{3}\sum_q \mathcal{L}_{q\bar{q}}(m^2)$ (see eq.~\eqref{eq:xsec}). The mass spectrum is divided by this quantity to counteract the fast decrease that it causes in the signal and bring it closer to a regularly oscillating function~\cite{Giudice:2017fmj}. For a general signal, the function $g(m)$ is defined such that the mass of the $n$th resonance is related to its index by $n=g(m_n)$. This is the quantity in terms of which the locations of the resonances are periodic and as such the quantity in terms of which the Fourier transform is best performed.\footnote{As the masses $m_n$ represent a discrete set and $g(m)$ must be a continuous function, simply requesting $g(m)$ to reproduce the correct masses does not fully define it and a smooth interpolation needs to be provided for the intermediate masses. In practice, there is usually an obvious definition for $g(m)$.} Simply doing a Fourier transform in terms of $m$ would not lead to an optimal significance when $g(m)$ differs too much from $m$. This is because a signal that varies over a wide range of frequencies over its duration would lead to a very wide peak in Fourier space. For CW/LD, we take $g(m)=R\sqrt{m^2 - k^2}$ (see eq.~\eqref{eq:masses})~\cite{Giudice:2017fmj}. Obviously, the power spectrum is expected to peak at $T=1$. The value of the peak can then be used as a test statistic.

\section{Comparison between the different methods}\label{Sec:Comparison}
\begin{figure}[t]
  \begin{center}
    \includegraphics[width=0.58\textwidth]{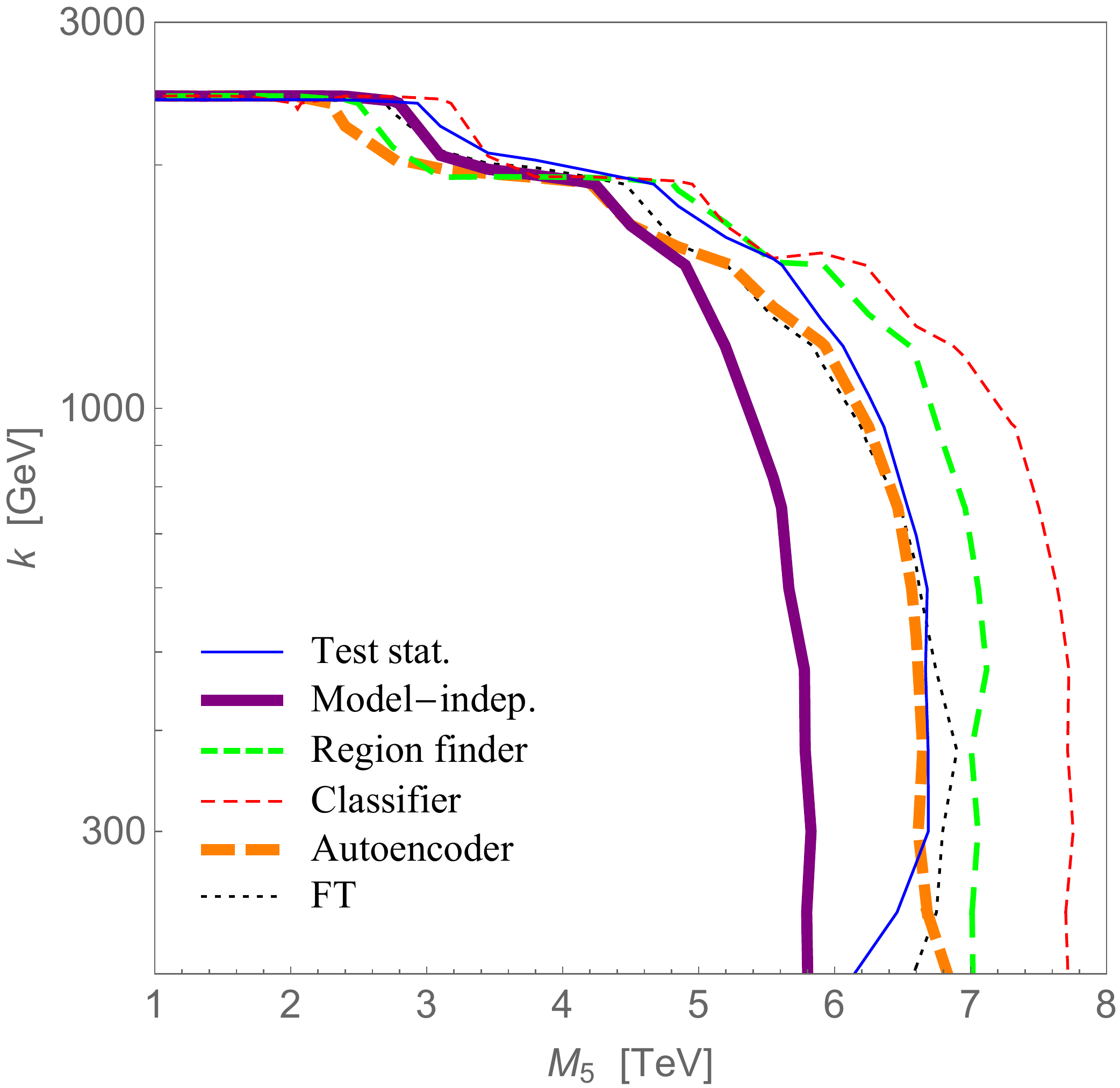}
  \end{center} 
  \caption{Comparison between the $2\sigma$ sensitivity of the various search methods, in the parameter space of the CW/LD model, based on the $\gamma\gamma$ dataset of ref.~\cite{Aaboud:2017yyg} (see appendix~\ref{app:App} for details). Thin lines correspond to methods designed for a specific model and specific parameters, medium lines to methods designed for a specific model but not a specific point and thick lines to searches for a general signal. Solid lines do not use machine learning and dashed ones do. The regions covered by the various methods are to the left of the corresponding curves.}\label{fig:Combined}
\end{figure}

To compare the different methods, we apply them to the CW/LD scenario with the $\gamma\gamma$ dataset of ref.~\cite{Aaboud:2017yyg}. The resulting reach in the parameter space of the model is shown in figure~\ref{fig:Combined}, where the contours correspond to a median expected significance of 2 sigma. As described in more detail in appendix~\ref{app:App}, the parameter $k$ is approximately the mass at which the spectrum begins, while $M_5$ controls the cross section, which is approximately proportional to $1/M_5^3$. The structure of the resonance masses $m_n$, if described in terms of $m_n/k$, is essentially independent of $k$ and $M_5$ in the range of parameters considered, and the asymptotic value of the mass splittings at high mass is given by $\Delta m \approx k/10$.

The blue (thin solid) curve corresponds to the test statistic method of section~\ref{sSec:WTSS}. The lower limit of the sum in eq.~(\ref{eq:TestStatSS}) was optimized to maximize the reach. A similar procedure was done for the upper limit, but we found that the results were virtually identical to those where the upper limit was taken as very large. In practice, we simply took $\min(10k, \text{2.7 TeV})$, where the latter is the upper limit of the experimental data. In terms of the reach in $M_5$, the sensitivity peaks at around $k \approx 600$~GeV. As $k$ moves upward, the limit on $M_5$ decreases as the total number of gravitons produced decreases. As $k$ moves downward, the part of the spectrum in which the splittings are resolvable experimentally also moves to lower masses, and the higher background at the lower masses lowers the sensitivity.

The purple (thick solid) curve corresponds to the model-independent search of section~\ref{sSec:GenericSignal}. Bins were considered as significant if their p-values were below 10$\%$. This value represents a fair compromise. A much lower value would reduce the sensitivity to weak signals, while for much higher values the selected regions would have such an extent that their interpretation would become unclear. As this approach covers a vast space of possible models, the bounds are unsurprisingly weaker than with the previous approach.

The green (medium thickness, dashed) curve corresponds to the anomalous region finder of section~\ref{sSec:RegionFinder}. The neural network was trained with signals with $k$ varying from 200 to 2700~GeV and $M_5$ varying from 1 to 6.5~TeV, where the latter number was obtained by optimization. The reach of this method is somewhat better than that of the first method thanks to the neural network's ability to assess whether a given excess is signal-like.

The red (thin dashed) curve corresponds to the classifier method of section~\ref{sSec:Classifier}. It gives the strongest bounds. One should note, however, that in this case one still needs to account for the look-elsewhere effect.

The orange (thick dashed) curve corresponds to the autoencoder of section~\ref{sSec:Autoencoders}. While the bounds are somewhat weaker than those of some of the other methods, one should note that the look-elsewhere effect is already taken into account in this case.

The black (thin dotted) curve corresponds to the bounds from Fourier analysis as in section~\ref{Sec:FourierAnalysis}. As in ref.~\cite{Giudice:2017fmj}, the lower limit of the integral was taken to be $m_{\text{min}}=k$, i.e.\ where the signal starts, and the upper limit $m_{\text{max}}$ was optimized at each point in the parameter space to maximize the expected significance. The bounds are mostly similar to those obtained using the CWT and a test statistic without machine learning.

It is also interesting to ask how the reach of the CWT-based methods compares with that of more traditional search strategies. Conveniently, CMS have performed a search for the same CW/LD scenario, based on a similar diphoton dataset, looking for a continuum excess at high masses~\cite{Sirunyan:2018wnk}. Their expected $95\%$ CL exclusion limits on $M_5$ were around $M_5 \approx 10$~TeV, which is comparable to the $2\sigma$ reach of $M_5 \approx 8$~TeV that we obtain here. One should keep in mind that the performance of the two approaches may compare differently when applied to other final states, to larger datasets, etc. Even more importantly, while the CW/LD example that we considered here does not present any special challenge to continuum excess searches such as ref.~\cite{Sirunyan:2018wnk}, other scenarios with oscillating signals can be elusive to such searches. As we discussed in the Introduction, this can happen due to negative contributions from destructive interference or due to the spectrum not extending to sufficiently high masses. Single-resonance searches provide expected limits of $M_5 \approx 5$~TeV, as obtained in ref.~\cite{Giudice:2017fmj} in the approximation that the sensitivity is dominated by the most prominent peak and that neighboring peaks do not hinder the search procedure. Obviously, higher sensitivity will be obtained by combining the contributions of multiple peaks, and the most natural way of implementing this in practice is via a Fourier or wavelet transform, as we do here.

\section{Discussion}\label{Sec:Conclusion}
The LHC experiments have by now developed very comprehensive sets of analyses that provide good coverage of essentially all the simple final states, as well as many exotic ones. They have discovered the Higgs boson, and progress is constantly being made on covering more and more of the parameter space in which physics beyond the Standard Model may be found. While no signs of new physics are seen yet, the theoretical expectation that at least the solution to the electroweak-Planck hierarchy problem is likely to be within the energy reach of the LHC calls for continuing the searches. However, as the LHC experiments have by now matured, improvements in the reach of existing techniques will mostly be gradual. It can therefore be very useful, as an alternative to just waiting, to think about new ways of looking at the data.

In this paper we proposed that wavelet transforms offer such a new way. It is a very general method that can be applied to many different final states at the LHC to search for periodic signals in kinematic distributions in a rather model-independent way. We have also pointed out examples of theoretical models, including some that address the hierarchy problem, for which such searches could be relevant.

We have designed and simulated five different approaches for processing the scalograms produced by the wavelet transforms. Part of these approaches use machine learning techniques, which is another direction into which new physics searches at the LHC can expand, as has been also proposed recently in several other contexts (e.g., refs.~\cite{Cohen:2017exh,Metodiev:2017vrx,Collins:2018epr,DAgnolo:2018cun,DeSimone:2018efk,Hajer:2018kqm,Heimel:2018mkt,Farina:2018fyg,Cerri:2018anq,Collins:2019jip,Blance:2019ibf}).

In our first approach, one assumes a specific new physics model and tests for the presence of its signature in the corresponding region of the scalogram using a simple test statistic. In the second, model-independent approach, the whole scalogram is being searched for extended regions of excess, and the most significant excess is assessed using the test statistic. In the third approach, the scalogram is being analyzed by a neutral network, which searches for regions of excess whose shape is consistent with excesses expected in a given class of models. The fourth approach tests for a specific signal using the classifier neural network, whose single output turns out to be a much more powerful test statistic than the more pedestrian test statistic of the first three methods. Finally, the fifth approach is a model-independent analysis in which an autoencoder neural network learns the background only, and identifies potential signals as deviations from a typical background.

We have exemplified the different methods and compared their sensitivities in the context of the diphoton invariant mass spectrum and the clockwork / linear dilaton model. We have also compared them with the Fourier transform method that was proposed in the context of the same model in ref.~\cite{Giudice:2017fmj}.

Our reach estimates should be viewed as conservative as there are various possible optimizations, either model-dependent or general, that we have left outside the scope of the current study. For example, we have not explored the possibility of using wavelets other than the Morlet wavelet. Also, we have done only very basic optimization of the architecture and the training parameters of the neural networks, so their performance is likely suboptimal. Nevertheless, all the methods resulted in sensitivities of up to $6$--$8$~TeV in $M_5$, which is comparable to existing searches that would be sensitive to the same scenario~\cite{Giudice:2017fmj,Sirunyan:2018wnk}. In addition, we would like to emphasize that similar analyses are interesting also in the dilepton~\cite{Aad:2019fac,Sirunyan:2018ipj}, dijet~\cite{ATLAS-CONF-2019-007,CMS-PAS-EXO-17-026} and other invariant mass spectra (and possibly other kinematic variables) and that the relative strengths and weaknesses of the different methods can vary depending on the final state, the new physics scenario, and the integrated luminosity. 

Wavelet-space searches can be of special importance if the new physics signal does not extend to sufficiently high masses where the background is low; if the new physics makes both positive and negative contributions to the mass spectrum (due to quantum interference); and in situations in which the systematic uncertainty on the normalization of the mass spectrum is a limiting factor. In these kinds of cases, signals can go undetected by existing experimental strategies (given a finite integrated luminosity), but discovered in wavelet-space searches. 

In conclusion, we hope to have convinced the reader that making wavelet-space searches part of the ATLAS and CMS toolkits is a possibility worth considering.

\noindent\textbf{Note added:} When this work was close to completion, ref.~\cite{Lillard:2019exp} appeared, which also proposed wavelet transforms as a way to search for new physics in kinematic distributions. Our approaches differ substantially due to the fact that ref.~\cite{Lillard:2019exp} uses the discrete wavelet transform based on the Haar wavelet, whereas our methods use the continuous wavelet transform.

\acknowledgments
We would like to thank Zvi Citron, Sanmay Ganguly, Noam Tal Hod, Marumi Kado, Enrique Kajomovitz, Jonathan Shlomi, and Margherita Spalla for useful discussions. HB is grateful to the Azrieli Foundation for the award of an Azrieli Fellowship. This research was supported in part by the Israel Science Foundation (grant no.\ 780/17).

\appendix

\section{Benchmark model: clockwork / linear dilaton (CW/LD)}\label{app:App}

In the \emph{linear dilaton} scenario, the Standard Model fields propagate on a brane in a space with one relatively large extra dimension. A particular scalar field, the dilaton, with a linear profile in the extra dimension, determines its warped geometry. The motivation for such a setup is its ability to explain the hierarchy between the electroweak and Planck scales. This scenario has first appeared in ref.~\cite{Antoniadis:2001sw}, inspired by the seven-dimensional gravitational dual~\cite{Aharony:1998ub,Giveon:1999px} of Little String Theory~\cite{Berkooz:1997cq,Seiberg:1997zk}. More recently, the same five-dimensional geometry has been rediscovered in~\cite{Giudice:2016yja} while exploring new applications for the \emph{clockwork mechanism}~\cite{Choi:2014rja,Choi:2015fiu,Kaplan:2015fuy}, in the limit of a large number of sites. Many phenomenological aspects of this scenario have been studied in refs.~\cite{Antoniadis:2011qw,Baryakhtar:2012wj,Cox:2012ee,Giudice:2017fmj}.

Most important for the collider phenomenology of the model are the Kaluza-Klein (KK) gravitons, whose masses (using the notation of ref.~\cite{Giudice:2017fmj}) are given by
\begin{equation}\label{eq:masses}
m_n^2 = k^2 + \frac{n^2}{R^2}\,,\qquad
n = 1, 2, 3, \ldots\,.
\end{equation}
The model parameters $k$ and $R$, which are related to the curvature and size of the extra dimension, are predicted to satisfy $kR \approx 10$ if this scenario is indeed responsible for the hierarchy. This implies a spectrum of narrowly-spaced resonances starting from mass $m_1 \simeq k$, with mass splittings that grow as a function of the mode number $n$ before reaching an asymptotic value of $\Delta m \simeq 1/R \approx k/10$ for $n \gg kR$. Near the beginning of the spectrum, the relative splittings $\Delta m/m$ are around a few percent (almost independent of the value of $k$), while their decrease as $1/m$ at large $n$ implies that at some point they fall below the experimental resolution. The intrinsic widths of the resonances are almost always negligible relative to the experimental resolution.  

The KK graviton fields $h_n^{\mu\nu}$ couple to the Standard Model stress-energy tensor $T_{\mu\nu}$ as
\begin{equation}
{\cal L} \supset -\frac{1}{\Lambda_n}\, h_n^{\mu\nu}\,T_{\mu\nu}\,,
\end{equation}
where
\begin{equation}
\Lambda_n^2 = M_5^3\pi R\left(1+\frac{k^2R^2}{n^2}\right)\,.
\end{equation}
Here $M_5$ is the five-dimensional reduced Planck mass, which is the fundamental scale of the theory. These couplings allow the KK gravitons to be produced from $gg$ and $q\bar q$ in $pp$ collisions with the cross sections
\begin{equation}\label{eq:xsec}
\sigma(pp \to G_n) = \frac{\pi}{48\Lambda_n^2}\left(3\,{\cal L}_{gg}(m_n^2) + 4\,\sum_q{\cal L}_{q\bar q}(m_n^2)\right),
\end{equation}
where
\begin{equation}
{\cal L}_{ij}(\hat s) = \frac{\hat s}{s}\, \int_{\hat s/s}^1 \frac{dx}{x}\, f_i(x)\, f_j\left(\frac{\hat s}{xs}\right)\,
\end{equation}
are the parton luminosities, for which we take the LO MSTW2008 PDFs~\cite{Martin:2009iq}. These couplings also allow the KK gravitons to decay to pairs of Standard Model particles, including $\gamma\gamma$. Heavy KK gravitons can also decay to pairs of lighter KK gravitons or KK scalars. We account for these decays in computing the $\gamma\gamma$ branching fraction, taking the case of rigid boundary conditions for the dilaton. However, since these decays start having an impact only for $m \gg k$, their effect is insignificant in the range of parameters we consider in this work. For additional details, see ref.~\cite{Giudice:2017fmj}. The $\gamma\gamma$ branching fraction ends up being about $4\%$, almost independent of the model parameters or the KK graviton mass.

Since the parameters $M_5$, $k$ and $R$ must combine to give the known value of the four-dimensional reduced Planck mass, $M_P \equiv 1/\sqrt{8\pi G}$, as $M_P^2 = (e^{2\pi kR} - 1)M_5^3/k$, only two of the parameters are independent, and we choose them to be $M_5$ and $k$. The parameter $k$ determines the beginning of the KK graviton spectrum, while $M_5$ fixes the cross section, which is approximately proportional to $1/M_5^3$.

We assume the experimental resolution in the diphoton invariant mass to be
\begin{equation}\label{eq:Resolution}
\frac{\sigma(m_{\gamma\gamma})}{m_{\gamma\gamma}} = \sqrt{\frac{a^2}{m_{\gamma\gamma}(\text{GeV})} + \frac{c^2}{2}}\,,
\end{equation}
with
\begin{equation}\label{eq:ResolutionParameters}
a = 12\%\,, \hspace{1cm} c = 1\%\,,
\end{equation}
which is based on partial information from refs.~\cite{ATLAS-CONF-2015-081, ATLAS-CONF-2016-059, Aaboud:2017yyg, Khachatryan:2016yec}. This resolution is used in figure~\ref{fig:Signal} and in the rest of the paper. We take the background from the search for heavy resonances decaying to two photons of ref.~\cite{Aaboud:2017yyg} (37~fb$^{-1}$ at 13~TeV), the ``Spin-2 selection''. A constant value of $\epsilon = 0.5$ is taken for the acceptance times efficiency.

\bibliography{biblio}
\bibliographystyle{utphys}

\end{document}